\documentclass[journal=jctcce]{achemso}
\setkeys{acs}{articletitle = true}
\pdfoutput=1

\usepackage[T1]{fontenc}
\usepackage[utf8]{inputenc}
\usepackage{color}
\usepackage{graphicx}
\usepackage{amsmath,amssymb}
\usepackage{calc}
\usepackage{feynmp}
\usepackage{bm}
\usepackage{subcaption}
\usepackage{multirow}
\usepackage{simplewick}
\usepackage{draftwatermark}
\usepackage[english]{babel}
\usepackage{multicol}
\usepackage{url}
\usepackage{hyperref}

\renewcommand{\vec}[1]{{\bm{#1}}}
\newcommand{\diagram}[2][0.5]{
  \raisebox{0.5ex-#1\height}{\includegraphics{#2}}
}

\newcommand{\tr}{{\operatorname{tr}}}

\newcommand{\Eq}[1]{Eq.~\eqref{#1}}

\newcommand{\xc}{ex\-change-cor\-re\-la\-tion}
\newcommand{\Hxc}{Har\-tree-\xc}
\newcommand{\Sc}{self-con\-sis\-tent}
\newcommand{\SC}{Self-con\-sis\-tent}

\title{%
  On the chemical potential
  of many-body perturbation theory
  in extended systems
}

\author{Felix Hummel}
\email{felix.hummel@tuwien.ac.at}
\affiliation{%
Institute for Theoretical Physics, TU Wien,\\
Wiedner Hauptstraße 8-10/136, 1040 Vienna, Austria}

\date{\today}
\SetWatermarkText{submitted manuscript}

\begin{document}

\setstretch{1.0}

\maketitle

\begin{abstract}
Many methods for computing electronic correlation effects
at finite temperature are related to many-body perturbation theory in the
grand-canonical ensemble. In most applications,
however, the average number of electrons is known rather than the
chemical potential, requiring that expensive correlation
calculations must be repeated iteratively in search for the
chemical potential that yields the desired average number of electrons.
In extended systems with mobile charges, however,
the long-ranged electrostatic interaction should guarantee
that the average ratio of negative and positive charges
is one for any finite chemical potential.
All properties per electron are virtually independent of the chemical potential,
as for instance in an electric wire at different voltage potentials.

This work shows that the infinite-size limit of the \xc\
free energy agrees with the infinite-size limit of the \xc\
grand potential at a non-interacting chemical potential. The latter requires
only one expensive correlation calculation for each system size.
Analogous to classical simulations of long-range-interacting particles,
this work uses a regularization of the Coulomb interaction such that each
electron on average interacts only with as many electrons as there are
electrons in the simulation, avoiding interactions with periodic images.

Numerical calculations of the warm uniform electron gas have been conducted
with the Spencer--Alavi regularization
employing the finite-temperature Hartree approximation for the
\Sc\ field and linearized finite-temperature direct-ring coupled
cluster doubles for treating correlation.
\end{abstract}

\begin{multicols}{2}

\section{Background}
In the warm-dense matter (WDM) regime
the relevant many-body states exceed the ground state
and the density is sufficiently large to require a quantum mechanical
treatment of the electrons interacting with each other.
WDM conditions are found, for instance,
during inertial confinement fusion (ICF),
in the core region of gas giants,
or in matter interacting with high intensity laser fields.\cite{graziani_2014}
Even at room temperature the thermal energy
must be considered to be large compared to the vanishing band gap
of bulk metals.

The mobility of electrons at warm-dense conditions
poses challenges for \emph{ab-initio} simulations of extended
systems that are absent in zero-temperature calculations.
Unlike at zero temperature, the number of electrons
in a volume of fixed shape fluctuates
rendering such a volume not necessarily charge neutral at all times.
Thus, the long-ranged Coulomb interaction cannot be used
under periodic boundary conditions due to the diverging electrostatic
energy per volume for net-charged configurations.
There are mainly two methods in current state of the art \emph{ab-initio}
simulations at warm-dense conditions to circumvent this divergence:
(i) The simulation is done in the canonical ensemble where
electrons are not permitted to enter or leave the simulated volume.
While this ensures charge neutrality it also reduces the number
of possible configurations, affecting the system's entropy.%
\cite{iyer_2015}
Path-integral quantum Monte Carlo (PIQMC) calculations are
usually conducted in the canonical ensemble.%
\cite{brown_2013,militzer_2019}
(ii)
Another possibility is to disregard the parts of
the electrostatic interaction stemming from the
average electron and background densities,
thus removing the divergence.
This allows for grand-canonical simulations with a fluctuating number
of electrons including its effect on the entropy.
Many-body perturbation theory calculations usually apply this method%
\cite{fetter_quantum_2003,thouless_2014}
following the work of Kohn and Luttinger, in particular the
assumption for arriving at
Eq.~(20) in Ref.~\citenum{kohn_ground-state_1960}.
A physical justification for this procedure would be
if the fluctuations of the positive background were fully
correlated with the fluctuations of the electrons.
Different mobilities of electrons and ions, however, question this
assumption.

In this work a third alternative is studied to treat
long-range electrostatic interactions with
thermal many-body perturbation theory.
Liang and coworkers\cite{Liang_2015}
have studied classical simulations of mobile electrostatically interacting
particles under periodic boundary conditions.
They look at the pair correlation function and
observe the theoretically expected Debey--Hueckel screening at long distances
only under two conditions:
(i) when simulating in the grand-canonical ensemble, and
(ii) when limiting the range of the electrostatic interaction, such
that the particles do not interact with all of their own periodic images.
Periodic boundary conditions cannot model charge fluctuations at
length scales beyond the size of the simulation cell.
In reality, the charges would move from one cell to the neighboring cell,
keeping the average charge constant.
Under periodic boundary conditions, however, charges can only appear or
disappear simultaneously in all periodic images of the simulation cell.
Still, the range of the electrostatic interaction can be limited to
allow for charge fluctuations.


A spherical truncation scheme has already
been developed by Spencer and Alavi\cite{spencer_2008} to prevent spurious
Fock-ex\-change interactions of the electrons with their periodic images
for zero-temperature calculations
as an alternative to other methods treating the occurring
integrable singularity.\cite{gygi_1986,carrier_2007}
Here, the truncation scheme is applied to all parts
of the electrostatic interaction
in the \Sc\ field calculations, as well as in
the subsequent perturbation calculation.
Other regularization schemes that limit the interaction range are also
possible, such as the Minimal Image Convention for atom centered
orbitals, or the Wigner--Seitz truncation scheme.\cite{%
irmler_2018,sundararaman_2013} For point-like charges the spherical
truncation is not continuous which may pose difficulties when considering
different atomic configurations.

\subsection*{Related work}
Finite-temperature many-body perturbation theory (FT-MBPT)
offers an elementary framework for ab-initio calculations of WDM.%
\cite{matsubara_new_1955,bloch_1958,bloch_1959,thouless_2014,fetter_quantum_2003}
Numerous approximation schemes employ thermal MBPT, such as
thermal second-order MBPT,%
\cite{nettelmann_2008,hirata_kohnluttinger_2013,son_2014,santra_2017}
finite-temperature random phase approximation,%
\cite{gupta_1980,perrot_1982,perrot_1984,csanak_1997}
Green's function based methods,%
\cite{vanLeeuwen_2006,welden_2016}
as well as some finite-temperature generalizations of coupled-cluster
methods.\cite{mandal_2003,white_2018,white_2020,hummel_2018}
An alternative formulation of the coupled-cluster methods has been
brought forward recently in the framework of thermo-field dynamics.%
\cite{harsha_2019a,harsha_2019b,harsha_2022}
Finite-temperature perturbation theory is originally formulated in the
grand-canonical ensemble, however formulations in the canonical ensemble exist.%
\cite{hirata_2019b,hirata_2020}
Equally, thermo field dynamics can be employed in the canonical ensemble.%
\cite{harsha_2020}

Analogous to ab-initio calculations at zero-temperature,
thermal Hartree--Fock and density functional theory (DFT) calculations
are among the most widely used methods.
\cite{mermin_1963,mermin_1965,pittalis_2011}
In general, it is not sufficient to use a zero-temperature \xc\ functional
and introduce temperature merely by smearing.
Temperature must be a parameter of the \xc\ functional.\cite{karasiev_2016}
At higher temperatures, a large number of one-body states is occupied with
non-negligible probabilities.
Orbital-free density functional theories (ofDFT)
aim at mitigating this with functionals that do not depend on the usual
Kohn--Sham orbital description of DFT.\cite{karasiev_2014,luo_2020}
Canonical or grand-canonical
full configuration interaction methods can be used for benchmarking
more approximate theories.\cite{hirata_2019a}
Finally, path-integral quantum Monte Carlo (PIQMC) methods are available
and often complement other calculations, as they have entirely different
error sources in the approximation of the many-body problem.
PIQMC calculations are usually conducted in the canonical ensemble.
\cite{brown_2013,militzer_2019}
High accuracy calculations of the warm uniform electron gas
are of particular interest since they can serve for accurate
temperature dependent parametrizations of DFT
\xc\ potentials.\cite{Sjostrom_2013,dornheim_2018,karasiev_2019}

The Kohn--Luttinger conundrum is also closely related to this work.
It states that the infinite-size zero-temperature limit of
finite-temperature many-body perturbation theory not necessarily agrees with
the infinite-size limit of zero-temperature many-body perturbation theory.
In the common approach where the zero-momentum part of the electrostatic
interaction is disregarded, certain terms called \emph{anomalous diagrams}
affect both, the chemical potential and the grand potential in a
way such that their contributions cancel in the zero-temperature limit of the
free energy under certain, but not all conditions.\cite{kohn_ground-state_1960}
Discussions on this conundrum can be found in Refs.\
\citenum{hirata_kohnluttinger_2013,son_2014,santra_2017,wellenhofer_2019,hirata_2022}.

With the method of this work the situation is different.
Considering the full electrostatic interaction with
a regularization in finite systems leads to a free energy per electron
that is asymptotically independent of the chemical potential
in the infinite-size limit. The electrostatic terms are strong
and do not allow a finite-order perturbative treatment, as discussed
in Subsection \ref{ssc:Scf} on fixed orbitals.
Although related to it, this work does not aim at solving the Kohn--Luttinger
conundrum. It may very well be that the long-ranged Coulomb interaction
causes a discontinuity at infinite-size and zero-temperature and
the result may depend on which limit is taken first.

\section{Methods}
\label{sec:WEG}
Let us now develop the regularization approach for the prototypical
warm-dense system: the warm uniform electron gas (UEG).
The UEG is a model of a metal, where
the positive ions of the lattice are replaced by a static homogeneous
positive background charge.
It has a vanishing band gap in the infinite-size limit
and thus qualifies for a warm-dense system at
all non-zero temperatures.
All properties of the warm UEG depend only on the thermodynamic state,
specified by its density and temperature.
The density is usually given in terms of the Wigner--Seitz radius $r_\mathrm s$
in atomic units,
such that the volume of a sphere with radius $r_\mathrm s$ corresponds to
the average volume per electron.
It is also convenient to specify the temperature in terms of
the dimensionless ratio $\Theta=k_\mathrm BT / \varepsilon_\mathrm F$,
where $k_\mathrm B T$ is the average thermal energy and
$\varepsilon_\mathrm F = k_\mathrm F^2/2$ is the Fermi energy
of a free non-spin-polarized, infinite
electron gas at the corresponding density and at zero temperature
with $k_\mathrm F^3 = 9\pi/4r_\mathrm s^3$.
This defines a natural temperature scale where different densities can
be compared to each other more directly.

The UEG is modeled by a finite cubic box of length $L$
under periodic boundary conditions having the volume
$\mathcal V = L^3 = 4\pi r_\mathrm{s}^3\mathcal N/3$.
It contains a homogeneous positive charge density with a total
charge of $\mathcal N$ elementary charges, which is considered fixed.
In the grand-canonical ensemble the number of electrons in the system
is not fixed but rather fluctuates around its expectation value
which depends on the chemical potential $\mu$.
Later, $\mu$ will be chosen such
that the expected number of electrons $N:=\langle\hat N\rangle$
equals, or is close to, the number of positive charges $\mathcal N$.
To treat the diverging electrostatic interaction
the Spencer--Alavi truncation of the
electrostatic interaction is used.
It is given by the usual Coulomb interaction
$1/r_{12}$ for the distance between two electronic coordinates
$r_{12}<\mathcal R$
and zero otherwise with the truncation radius
$\mathcal R = r_\mathrm s\mathcal N^{1/3}$.
The kernel of this interaction within a sum over momenta is
$V(q) = 4\pi(1- \cos q\mathcal R)/\mathcal Vq^2$.
For finite $\mathcal N$ the kernel is also finite at $q=0$
and evaluates to $2\pi \mathcal R^2/\mathcal V$.
With this choice
the interaction ``sees'' on average $\mathcal N$ electrons and
it reduces to the usual electrostatic interaction in the limit
$\mathcal N\to\infty$.\cite{spencer_2008}

All states are expanded in anti-symmetrized
products of one-electron wavefunctions
that are eigenfunctions of the single-electron
kinetic operator $-\vec{\nabla}^2/2$ under periodic boundary conditions.
The normalized eigenfunctions are the plane waves commensurate with
the box length
\begin{equation}
  \psi_{\vec k\sigma}(\vec r,\tau)
  = \frac{1}{\sqrt \mathcal V} e^{-i\vec k\cdot \vec r} \delta_{\sigma\tau},
\end{equation}
with $\vec k\in 2\pi\mathbb Z^3/L$ and
where $\sigma,\tau\in\{\uparrow,\downarrow\}$
denote the spin coordinate of the wavefunction and the electron, respectively.
With the operator $\hat c^\dagger_{\vec k,\sigma}$ creating an electron in
the state $\vec k,\sigma$ and $\hat c_{\vec k,\sigma}$ annihilating it,
the electronic Hamiltonian
of the modeled UEG reads
\begin{align}
  \hat H &= \hat T + \hat V_\mathrm{ext} + \hat V \nonumber\\\nonumber
  &= \sum_{\vec k,\sigma} \frac{\vec k^2}2
    \hat c_{\vec k,\sigma}^\dagger \hat c_{\vec k,\sigma}
  - \sum_{\vec k,\sigma} V(0)\,\mathcal N\,
    \hat c_{\vec k,\sigma}^\dagger \hat c_{\vec k,\sigma} \\
  &+ \frac12\sum_{\vec k,\sigma,\vec k',\sigma',{\vec q}}
    V(|\vec q|)\,
    \hat c_{\vec k+{\vec q},\sigma}^\dagger \hat c_{\vec k'-{\vec q},\sigma'}^\dagger
    \hat c_{\vec k',\sigma'} \hat c_{\vec k,\sigma}.
  \label{eqn:HUeg}
\end{align}
It consists of three terms: the kinetic term $\hat T$,
the electron--background interaction $\hat V_\mathrm{ext}$ and
the electron--electron interaction $\hat V$, respectively.
Exact diagonalization of the Hamiltonian is infeasible except for very
limited system sizes.
This work shall also employ the approximation approach
of computational materials science, where one first
performs a \Sc\ field (SCF) calculation, followed by
a perturbative approximation of the correlation based on the SCF result.
In accordance with the common workflow of Random Phase Approximation (RPA)
calculations for low-band-gap systems, the SCF only employs the
Hartree approximation rather than Hartree--Fock and
exchange is considered at first order non-\Sc ly.%
\cite{harl_assessing_2010}
The finite temperature correlation contributions are estimated by a
linearized form of the direct ring coupled cluster doubles approximation.

\subsection{\SC\ field in the Hartree approximation}
\label{ssc:Scf}
In the \Sc\ field approach the
two-body operator in the electron--electron interaction $\hat V$ is
partially contracted to
a one-body interaction.\cite{mermin_1963}
In the Hartree approximation only the direct contraction is considered and
the resulting one-body operator is given by
\begin{equation}
  \hat H_0 = \hat T + \hat V_\mathrm{ext}
  + \sum_{\vec k,\sigma,\vec k',\sigma'}
    V(0)\,
    \hat c_{\vec k,\sigma}^\dagger \hat c_{\vec k,\sigma}
    \langle \hat c_{\vec k',\sigma'}^\dagger
    \hat c_{\vec k',\sigma'}\rangle_0
  \label{eqn:H0}
\end{equation}
where $\langle\hat A\rangle_0$ denotes the one-body thermal
equilibrium expectation value of the operator $\hat A$, defined by
\begin{equation}
  \langle \hat A\rangle_0 = \frac{\tr\{\hat A \hat\rho_0\}}{\tr\{\hat\rho_0\}}
\end{equation}
with the (non-normalized) one-body density matrix
\begin{equation}
  \hat\rho_0 = \exp\{-\beta(\hat H_0 - \mu\hat N)\}.
  \label{eqn:rho0}
\end{equation}
All terms in \Eq{eqn:H0} are diagonal in the chosen basis so we can immediately
write the equations for the eigenvalue of each state $i=(\vec k_i,\sigma_i)$
\begin{equation}
  \label{eqn:ScfEigen}
  \varepsilon_i = \frac{\vec k_i^2}2 + V(0)
  \left(
    N_0 - \mathcal N
  \right)
\end{equation}
with $N_0
  = \sum_i 1/(e^{\beta(\varepsilon_i-\mu)}+1)$,
Introducing the notation
$\varepsilon_i = \vec k_i^2/2 + \Delta\varepsilon$,
we have to find a shift of
eigenenergies $\Delta\varepsilon$, uniform for all states,
satisfying the non-linear equation
\begin{equation}
  \Delta\varepsilon = V(0)
  \underbrace{
  \left(
    \sum_i\frac1{e^{\beta(\vec k_i^2/2+\Delta\varepsilon-\mu)}+1}
    - \mathcal N
  \right)
  }_{=:\Delta N_0}
  \label{eqn:DeltaEps}
\end{equation}
for the given thermodynamic state point $(\mu,\mathcal V,\beta)$.
So far, the number of positive charges $\mathcal N$ is an independent parameter.
The quantity $\Delta N_0$ denotes the net-negative charge of the system
in the non-interacting approximation.
Note that it may differ from zero for charge neutral systems as
the fully interacting $N=\langle\hat N\rangle$ may differ from its
non-interacting approximation $N_0=\langle\hat N\rangle_0$.
Having solved \Eq{eqn:DeltaEps} for $\Delta\varepsilon$ we can evaluate
the non-interacting grand potential
\begin{equation}
  \Omega_0 = -\frac1\beta \sum_i
    \log\left(1+e^{-\beta(\vec k_i^2/2 - \eta)}\right)
\end{equation}
with the \emph{effective chemical potential} $\eta:=\mu-\Delta\varepsilon$.
If we want to compare energies per electron for different system sizes
we also need to account for the background--background interaction energy,
which is independent of the electronic degrees of freedom.
Furthermore, pairwise interactions in $\Omega_0$ are double-counted in the SCF.
Accounting for both contributions yields the mean-field grand potential
in the Hartree approximation
\begin{equation}
  \Omega_\mathrm{H} = \Omega_0
  + \frac12\, V(0)\left(
    \mathcal N^2 - N_0^2
  \right).
  \label{eqn:OmegaH}
\end{equation}
Note that the expected number of electrons in the Hartree approximation
$N_\mathrm{H}=-\partial_\mu\Omega_\mathrm{H}$
equals the expected number of electrons of the non-interacting
system $N_0=-\partial_\mu\Omega_0$, which is
given by the sum $\sum_i n_i$ of non-interacting occupancies
$n_i
  = \langle \hat c_i^\dagger \hat c_i\rangle_0
  = 1/(e^{\beta({\vec k_i}^2/2 - \eta)}+1)$.

The Hartree grand potential $\Omega_\mathrm{H}(\mu)$ is a function
of the chemical potential $\mu$. Of particular interest is
the chemical potential $\mu_\mathrm{H}$ for which the expected number of
electrons in the Hartree approximation $N_\mathrm{H}$ matches the number of
positive charges $\mathcal N$. This chemical potential is chosen for
determining the Hartree free energy from a Legendre transformation
$F_\mathrm{H}=\Omega_\mathrm{H}(\mu_\mathrm{H}) + \mu_\mathrm{H}\mathcal N$.
For this particular chemical potential, the solution of the Hartree equation
$\Delta\varepsilon=0$ follows trivially from \Eq{eqn:DeltaEps} and
the effective chemical potential
$\eta_\mathrm{H}=\mu_\mathrm{H}-\Delta\varepsilon$ equals the Hartree
chemical potential.

\subsubsection*{A rough estimate of the Hartree \Sc\ field}
Before turning to the other contributions to the grand potential,
it interesting to estimate the Hartree solution $\Delta\varepsilon$
for chemical potentials $\mu$ close to the
chemical potential $\mu_\mathrm{H}$,
which satisfies $N_\mathrm{H}=\mathcal N$.
To this end, we expand the expected number of electrons
$N_\mathrm{H}=-\partial_\mu\Omega_0$
as a function of the effective chemical potential $\eta=\mu-\Delta\varepsilon$
at the Hartree effective chemical potential $\eta_\mathrm{H}=\mu_\mathrm{H}$,
where $\Delta\varepsilon=0$. The expansion reads
\begin{multline}
  N_\mathrm{H}(\eta)
  = \mathcal N
  - (\eta-\mu_\mathrm{H})\partial^2_{\eta\mu}\Omega_0(\mu_\mathrm{H}) \\
  + \mathcal O\left(
    (\eta - \mu_\mathrm{H})^2
  \right)
\end{multline}
where $\partial^2_{\eta\mu}\Omega_0(\mu_H) = -\beta\sum_i n_i^i(\mu_\mathrm{H})$
with the shorthand notation $n_i^i=n_i(1-n_i)$.
We can approximate the difference $\Delta N_0=N_\mathrm{H}-\mathcal N$
between the expected number of
electrons in the SCF calculation and the number of positive charges
to first order in $(\eta-\mu_\mathrm{H})$ by
\begin{equation}
\label{eqn:ScfNumberExpectation}
  \Delta N_0
  \approx
    -(\mu - \Delta\varepsilon - \mu_\mathrm{H})
    \partial^2_{\eta\mu}\Omega_0(\mu_\mathrm{H})
\end{equation}
We further assume that
in the warm-dense-matter regime
the term $\sum_i n_i(1-n_i)$
scales linearly with system size.
Together with \Eq{eqn:DeltaEps} $\Delta \varepsilon = V(0)\Delta N_0$
we are now in the position to approximately
solve \Eq{eqn:ScfNumberExpectation} for $\Delta N_0$:
\begin{equation}
  \Delta N_0 \approx \frac{
    \partial^2_{\eta\mu}\Omega_0
  }{
    V(0)\partial^2_{\eta\mu}\Omega_0 - 1
  }\,(\mu - \mu_\mathrm{H}).
\end{equation}
Inserting $\Delta N_0$ into \Eq{eqn:DeltaEps}
and expanding in powers of $\mathcal N$ for large $\mathcal N$ finally yields
\begin{align}
  \label{eqn:DeltaEpsLargeN}
  \nonumber
\Delta\varepsilon \approx &\, (\mu-\mu_\mathrm{H}) \\
  &\left(
  1 + \frac{2r_\mathrm{s}}{3\partial^2_{\eta\mu}\Omega_0}\, \mathcal N^{\frac13}
    + \mathcal O(\mathcal N^{-\frac43})
  \right), \\
  \label{eqn:DeltaNLargeN}
\Delta N_0 \approx &\, (\mu-\mu_\mathrm{H})
  \,\frac{2r_\mathrm{s}}3\, \mathcal N^{\frac13}
    + \mathcal O(\mathcal N^{-\frac13}) \,
\end{align}
where we inserted $V(0) = 2\pi \mathcal R^2/\mathcal V$.

This is an essential result.
When changing the chemical potential, the relaxed \Sc\ field
eigenenergies asymptotically change in the exact same way for
large enough system sizes.
The numerical results in Subsection \ref{ssc:ResultsH} indicate
that this behavior already sets in already for relatively low system sizes.
Since only the difference of the chemical potential $\mu$ and
the eigenenergies $\varepsilon_{i}=\vec k_i^2/2 + \Delta\varepsilon$
enter in subsequent calculations, size-intensive observables will
be asymptotically independent of the choice of $\mu$ in
the thermodynamic limit.
At warm-dense conditions, any choice is acceptable, including $\mu=0$,
which is the classical definition of the chemical potential
of electrons in a grounded conductor that can supply or absorb
any number of electrons.

\subsubsection*{Using fixed instead of relaxed orbitals}
If one chooses to work with a fixed set of orbitals and eigenenergies
for various values of the chemical potential $\mu$,
the electron--background interaction $\diagram{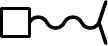}$ and the
electron--electron density interaction $\diagram{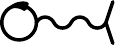}$ do not cancel
for $\mu$ deviating from $\mu_\mathrm{H}$.
There are terms of order $n$ in the perturbation series
whose contribution scales as $\mathcal O(\mathcal N(-\mathcal N^{2/3})^n)$,
which is super-extensive for each $n$ and alternating in sign.
This can be cured by summing
the interactions to infinite order, which is equivalent to performing
an SCF calculation at the modified chemical potential.\cite{mattuck_1992}

\subsubsection*{Fluctuations of the number of electrons}
From $N_\mathrm{H} = \mathcal N + \Delta N_0$ we can
also estimate the variance $\delta N_\mathrm{H}^2$ of the fluctuations of the number
of electrons in the SCF calculation for large $\mathcal N$ by
\begin{equation}
\label{eqn:ScfVariance}
  \delta N_\mathrm{H}^2 = \frac1\beta\frac{\partial N_\mathrm{H}}{\partial\mu}
  \approx \frac{2r_\mathrm{s}}{3\beta} \mathcal N^{\frac13}
  + \mathcal O\left(\mathcal N^{-\frac13}\right).
\end{equation}
This agrees qualitatively with classical charge fluctuations $\delta N^2_\mathrm{C}$
on the surface of a
grounded conducting sphere of radius
$\mathcal R=r_\mathrm s \mathcal N^{1/3}$,
found from the equipartition theorem
\begin{equation}
  \frac{\delta N_\mathrm{C}^2}{\mathcal R} = \frac1\beta\,.
\end{equation}
Note that we need to consider the response of the one-body energies
to changes of the chemical potential for computing derivatives of the
grand potential beyond first order, such as
$\partial N_\mathrm{H}/\partial\mu = -\partial^2\Omega_0/\partial\mu^2$
in \Eq{eqn:ScfVariance}.
For comparison, a fixed density matrix $\hat\rho_0$
that separates into a product of one-body density matrices
is only capable of describing electron-number fluctuations of the form
$\delta N^2_0=\sum_in_i(1-n_i)$, which
is proportional to $\mathcal N$ rather than to $\mathcal N^{1/3}$
at warm-dense conditions.

\subsection{First-order exchange}
The \Sc\ field approximation is crude but computationally
efficient. To improve on the approximation,
finite temperature many-body perturbation theory (FT-MBPT)
offers an expansion of the grand potential
in powers of the difference $\hat H_1 = \hat H - \hat H_0$
between the true Hamiltonian $\hat H$ and the \Sc\ field
Hamiltonian $\hat H_0$.
Having employed the Hartree approximation for the SCF, the leading order
term is the first-order exchange term
\begin{equation}
  \diagram{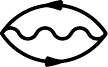}
  = -\frac12 \sum_{ij} n_{ij} V^{ij}_{ji},
  \label{eqn:FirstOrder}
\end{equation}
where we again use compound indices $i=({\vec k}_i,\sigma_i)$ and
$j=({\vec k}_j,\sigma_j)$
to denote the spatial and spin components of the respective spin-orbitals.
We also employ the shorthand notation $n_{ij\ldots}=n_in_j\ldots$ for
products of one-body occupancies.
$V^{pq}_{sr}$ denotes the components of the electron--electron
interaction operator in the basis of the plane-wave spin-orbitals such that
$
  \hat V = \frac12 \sum_{pqrs} V^{pq}_{sr}\,
    \hat c_{{\vec k}_p,\sigma_p}^\dagger \hat c_{{\vec k}_q,\sigma_q}^\dagger
    \hat c_{{\vec k}_r,\sigma_r} \hat c_{{\vec k}_s,\sigma_s}.
$
For a translationally invariant, isotropic interaction the components read
\begin{equation}
  V^{pq}_{sr} =
    \delta^{\sigma_p}_{\sigma_s}\,\delta^{\sigma_q}_{\sigma_r}\,
    \delta_{{\vec k}_p+{\vec k}_q-{\vec k}_r-{\vec k}_s}\,
    V\left(|{\vec k}_q-{\vec k}_r|\right).
\end{equation}
The first-order exchange term with non-Hartree--Fock orbitals is often
referred to as \emph{exact exchange} (EE). It is given by
\begin{equation}
\label{eqn:OmegaX}
  \Omega_\mathrm{x} =
    - \frac12 \sum_{ij} n_{ij}\,
  \delta^{\sigma_i}_{\sigma_j}V\left(|{\vec k_i}-{\vec k_j}|\right).
\end{equation}
Adding the first-order exchange contribution $\Omega_\mathrm{x}$
to $\Omega_\mathrm{H}$
yields the improved Hartree-ex\-change approximation $\Omega_\mathrm{Hx}$.

\subsection{Linearized direct-ring coupled cluster}
Let us now turn to correlation and exchange effects beyond first order.
Here, it is treated at the level of
linearized direct-ring coupled-cluster doubles (ldrCCD) theory.%
\cite{hummel_2018}
A truncation of the perturbation expansion at any finite order
diverges for the uniform electron
gas in the zero-temperature and infinite-size limit.
However, summing over the so-called ring terms up to infinite order
yields convergent results in that limit.%
\cite{macke_uber_1950,pines_collective_1952}
Although finite-order expansions always converge at finite temperature,
we desire a theory with a uniform convergence behavior
for $T\rightarrow 0$, at least in principle.
ldrCCD is one of the simplest theories providing
this resummation of the ring terms.
It contains all ring terms that can be formed
with exactly two particle/hole pairs and additionally contains
their corresponding screened-ex\-change terms.
It is determined by the finite-temperature linearized direct-ring
coupled-cluster amplitude integral equations
\begin{multline}
  T^{ab}_{ij}({\tau})
  = (-1)\int_0^\tau d\tau'\,
  e^{-({\tau}-{\tau'}){\Delta}^{ab}_{ij}}
  \Bigg[ V^{ab}_{ij} \\
  + \sum_{ck}n^c_k V^{kb}_{cj}T^{ac}_{ik}({\tau'})
  + \sum_{dl}n^d_l V^{al}_{id}T^{db}_{lj}({\tau'}) \Bigg]
  \label{eqn:ldrCcdAmplitudes}
\end{multline}
with $\Delta^{ab}_{ij}=\varepsilon_a-\varepsilon_i+\varepsilon_b-\varepsilon_j$
and where we now also need products of vacancy and occupancy
probabilities, denoted by $n^c_k=(1-n_c)n_k$.
\Eq{eqn:ldrCcdAmplitudes} can also be given in terms of diagrams
\begin{equation*}
  \diagram[1]{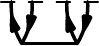}
  =
  \hspace*{-2ex}
  \diagram[1]{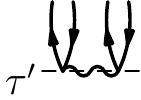}
  + \diagram[1]{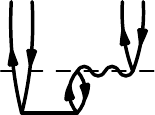}
  + \diagram[1]{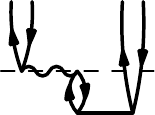}\,.
\end{equation*}
With the solutions of the
amplitude functions $T^{ab}_{ij}(\tau)$, satisfying \Eq{eqn:ldrCcdAmplitudes}
on the interval $\tau\in[0,\beta]$ with the initial conditions
$T^{ab}_{ij}(0)=0$,
the ldrCCD grand potential can be evaluated from
\begin{multline}
  \Omega_\mathrm{c}
  = \diagram{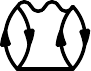}
  + \diagram{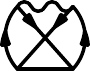} \\
  = \frac1\beta\int_0^\beta {d\tau}\,\Bigg[
    \frac12\sum_{abij}n^{ab}_{ij} \left(V^{ij}_{ab}-V^{ji}_{ab}\right)
    T^{ab}_{ij}(\tau)\Bigg]
  \label{eqn:CcdEnergy}
\end{multline}
with $n^{ab}_{ij}=(1-n_a)n_i(1-n_b)n_j$.
All indices iterate in principle over the infinite number of plane wave states.
Practical truncation schemes are discussed in Section \ref{ssc:ldrccdResults}.
The linear system of coupled integral equations in \Eq{eqn:ldrCcdAmplitudes}
can be solved
by diagonalizing an effective particle/hole interaction $\tilde H$, analogous to
the Tamm--Dancoff approximation of the Casida equations at zero temperature.
The effective particle/hole interaction reads
\begin{equation}
  \label{eqn:effectiveH}
  \tilde H^{bi}_{ja} =
    \delta^b_a\delta^i_j \Delta^b_j + \sqrt{n^{ab}_{ij}}\, V^{bi}_{ja}
  = U^b_{jF} \Lambda^F_F {U^\ast}^{iF}_a,
\end{equation}
which, interpreting the indices $(b,j)$ as a compound row index
and the indices $(a,i)$ as a compound column index, is a
hermitian matrix and thus permits a real-valued
eigendecomposition.
We can then transform the electron repulsion integrals with and without
exchange into the space of eigenmodes
\begin{align}
  \label{eqn:EigenModeW}
  W_{FG} &=
    \sum_{abij} U^a_{iF} U^b_{jG} \sqrt{n^{ab}_{ij}}\,
    (V^{ij}_{ab} - V^{ji}_{ab}), \\
  \label{eqn:EigenModeV}
  V_{FG} &=
    \sum_{abij} U^a_{iF} U^b_{jG} \sqrt{n^{ab}_{ij}}\, V^{ij}_{ab},
\end{align}
and finally retrieve the ldrCCD approximation of the
correlation grand potential from
\begin{multline}
\label{eqn:OmegaC}
\Omega_\mathrm{c} = \\
  -\sum_{FG}\left(
  \frac1{\Lambda_{FG}} + \frac{e^{-\beta\Lambda_{FG}}-1}{\beta\Lambda_{FG}^2}
  \right) \frac12\, W_{FG} {V^\ast}^{FG}
\end{multline}
with $\Lambda_{FG} = \Lambda^F_F + \Lambda^G_G$ and
${V^\ast}^{FG} = \overline{V_{FG}}$ denoting the conjugate transpose.%
\cite{hummel_2018}

\subsection{Free energies}
So far, we have discussed all considered contributions
to the grand potential
\begin{equation}
  \label{eqn:OmegaHxc}
  \Omega_\mathrm{Hxc}(\mu) =
  \Omega_\mathrm{H}(\mu)  + \Omega_\mathrm{x}(\mu) + \Omega_\mathrm{c}(\mu)
\end{equation}
as
a function of the thermodynamic state point
in the grand-canonical ensemble, in particular of the chemical potential $\mu$.
The number of positive charges $\mathcal N$ is merely a system parameter.
We are, however, interested
in the free energy $F_\mathrm{Hxc}(\mathcal N)$ of the
charge-neutral system where the expected number of electrons
$N_\mathrm{Hxc} := -\partial_\mu\Omega_\mathrm{Hxc}$ equals the fixed number
$\mathcal N$ of positive charges.
It is found from the Legendre transformation
\begin{equation}
  \label{eqn:HxcFreeEnergy}
  F_\mathrm{Hxc}(\mathcal N) =
    \Omega_\mathrm{Hxc}(\mu_\mathrm{Hxc})
    + \mu_\mathrm{Hxc} \mathcal N
\end{equation}
where $\mu_\mathrm{Hxc}$ satisfies the charge-neutrality condition
for the \Hxc\ grand potential
$-\partial_\mu \Omega_\mathrm{Hxc}(\mu_\mathrm{Hxc})
  = \mathcal N$.
The final quantity of interest is the
\emph{\xc\ (xc) free energy}
$F_\mathrm{xc}
  = F_\mathrm{Hxc} - F_\mathrm{H}$
beyond the free energy
of the \Sc\ field solution
$F_\mathrm{H}(\mathcal N)
  = \Omega_\mathrm{H}(\mu_\mathrm{H}) + \mu_\mathrm{H}\mathcal N$,
where $\mu_\mathrm{H}$ satisfies the charge-neutrality
condition for the Hartree grand potential
$-\partial_\mu \Omega_\mathrm{H}(\mu_\mathrm{H}) = \mathcal N$.
Note, that in general the \Hxc\ chemical potential
$\mu_\mathrm{Hxc}$ differs from the Hartree chemical potential $\mu_\mathrm{H}$,
which is the non-interacting chemical potential.

\subsubsection*{A rough estimate of the
ex\-change-cor\-re\-la\-tion free energy}
Let us now estimate the behavior of $\mu_\mathrm{Hxc}$ and $F_\mathrm{xc}$
for large system sizes $\mathcal N$.
We start by looking at the charge-neutrality condition
$-\partial_\mu \Omega_\mathrm{Hxc}(\mu_\mathrm{Hxc}) = \mathcal N$,
for the \Hxc\
chemical potential $\mu_\mathrm{Hxc}$.
From \Eq{eqn:OmegaHxc} we can immediately write the expected number of
electrons as $-\partial_\mu\Omega_\mathrm{H} - \partial_\mu\Omega_\mathrm{xc}$,
where $-\partial_\mu\Omega_\mathrm{H} = N_\mathrm{H}$ is the expected
number of electrons in the Hartree approximation at the interacting
chemical potential $\mu_\mathrm{Hxc}$, which differs from $\mathcal N$ for
$\mu\neq\mu_\mathrm{H}$.
At the end of Subsection \ref{ssc:Scf} we have estimated that it
behaves as
$\mathcal N
  + (\mu_\mathrm{Hxc}-\mu_\mathrm{H})r_\mathrm{s}\,\mathcal O(\mathcal N^{1/3})$
for sufficiently large $\mathcal N$, according to \Eq{eqn:DeltaNLargeN}.
From Eqs.~\eqref{eqn:OmegaX} and \eqref{eqn:OmegaC} it follows that
the only terms that depend on $\mu$ in
the remaining contribution $\Omega_\mathrm{xc}$
are the occupancy and vacancy expectation values
$n_i=1/(e^{\beta(\varepsilon_i - \mu)}+1)$ and $n^a=1-n_a$,
respectively.
The expectation values $n_i$ depend only on the difference
$\varepsilon_i-\mu$ between the eigenenergies
$\varepsilon_i=\vec k_i^2/2 + \Delta\varepsilon$
and the chemical potential $\mu$, where $\Delta\varepsilon$ is
the shift of eigenenergies, uniform for all states $i$,
found from the \Sc\ field
solution for the interacting chemical potential $\mu_\mathrm{Hxc}$.
Using the notion of the effective chemical potential
$\eta = \mu-\Delta\varepsilon$ introduced in Subsection \ref{ssc:Scf},
we can write the derivative with respect to the chemical potential
in terms of a derivative with respect to the effective chemical potential
from the chain rule
\begin{equation}
  \label{eqn:DOmegaDMu}
  -\partial_\mu\Omega_\mathrm{xc}(\mu) =
  -(\partial_\eta \Omega_\mathrm{xc})
  (\partial_\mu \eta(\mu)).
\end{equation}
We have already estimated the asymptotic behavior of $\Delta\varepsilon$
in \Eq{eqn:DeltaEpsLargeN} from which we can find the behavior of
$\partial_\mu\eta(\mu)$ for large $\mathcal N$:
\begin{equation}
  \label{eqn:deta_dmu}
  \partial_\mu\eta(\mu)
  \approx -\frac{2r_\mathrm{s}}{3\partial^2_{\eta\mu}\Omega_0}\, \mathcal N^{\frac13}
  + \mathcal O(\mathcal N^{-\frac43}).
\end{equation}
Note that $-\partial^2_{\eta\mu}\Omega_0=\beta\sum_i n_i^i$
scales linearly with $\mathcal N$ under warm-dense conditions.
Similarly, since $\partial_\eta n^a = -\beta n_a^a$, we can also assume that
$-\partial_\eta \Omega_\mathrm{xc}$ scales at most linearly with
the system size $\mathcal N$ under these conditions.
Collecting all contributions to the expected number of electrons
gives
\begin{multline}
  -\partial_\mu\Omega_\mathrm{Hxc}(\mu_\mathrm{Hxc}) \approx \mathcal N \\
  + \left(
    \mu_\mathrm{Hxc}-\mu_\mathrm{H}
    + \frac{\partial_\eta\Omega_\mathrm{xc}}{\partial^2_{\eta\mu}\Omega_0}
  \right) \frac{2r_\mathrm{s}}3\,\mathcal N^{\frac13}
  + \mathcal O(\mathcal N^{-\frac13})
\end{multline}
where the fraction inside the parentheses does not depend on $\mathcal N$
asymptotically.
Remarkably, this means
that the expected number of electrons per positive
charge $N_\mathrm{Hxc}/\mathcal N \approx 1+ \mathcal O(\mathcal N^{-2/3})$
converges asymptotically to one for large system sizes for any choice
of the chemical potential $\mu_\mathrm{Hxc}$.
Still, the absolute
deviation of $N_\mathrm{Hxc}$ from $\mathcal N$ does depend on
$\mu_\mathrm{Hxc}$ and scales as
$\mathcal O(\mathcal N^{1/3})$ with the number of positive charges
$\mathcal N$.
From this deviation we can approximately solve the charge-neutrality condition
to find the \Hxc\ chemical potential:
\begin{equation}
  \label{eqn:MuHxcLargeN}
  \mu_\mathrm{Hxc}-\mu_\mathrm{H}
    \approx - \frac{
      \partial_\eta\Omega_\mathrm{xc}(\mu_\mathrm{H})
    }{
      \partial^2_{\eta\mu}\Omega_0(\mu_\mathrm{H})
    },
\end{equation}
Although the expected number of electrons per positive charge converges
to one for any chemical potential in the thermodynamic limit,
there is a non-vanishing deviation from the non-interacting
chemical potential $\mu_\mathrm{H}$
required if also the absolute expected number of electrons $N_\mathrm{Hxc}$
should match the number of positive charges for large $\mathcal N$.

Knowing the asymptotic behavior of the interacting chemical potential
$\mu_\mathrm{Hxc}$ we can now estimate the free energy for large system
sizes. For that purpose, we expand the \Hxc\ free energy at the
non-interacting chemical potential in \Eq{eqn:HxcFreeEnergy} in powers
of the difference $(\mu_\mathrm{Hxc}-\mu_\mathrm{H})$, which we have
found to be finite but approximately independent of $\mathcal N$:
\begin{multline}
  F_\mathrm{Hxc} =
    \Omega_\mathrm{Hxc}(\mu_\mathrm{H})
    + (\mu_\mathrm{Hxc}-\mu_\mathrm{H})
      \partial_\mu\Omega_\mathrm{Hxc}(\mu_\mathrm{H}) \\
    + \mathcal O\left((\mu_\mathrm{Hxc}-\mu_\mathrm{H})^2\right)
    + \mu_\mathrm{Hxc} \mathcal N
\end{multline}
Subtracting the Hartree free energy
$F_\mathrm{H}=\Omega_\mathrm{H}(\mu_\mathrm{H}) + \mu_\mathrm{H}\mathcal N$
we arrive at an estimate of the \xc\ free energy expansion
\begin{multline}
  \label{eqn:FxcExpansion}
  F_\mathrm{xc} \approx \Omega_\mathrm{xc}(\mu_\mathrm{H})
  + (\mu_\mathrm{Hxc}-\mu_\mathrm{H})
  \partial_\mu \Omega_\mathrm{xc}(\mu_\mathrm{H}) \\
    + \mathcal O\left((\mu_\mathrm{Hxc}-\mu_\mathrm{H})^2\right).
\end{multline}
Our estimate of $\partial_\mu\eta$ in \Eq{eqn:deta_dmu} is approximately
independent of $\mu$. Therefore,
the higher derivatives of the \xc\ grand potential occurring in the above
expansion
are estimated to be of the form
$\partial^n_{\mu\ldots}\Omega_\mathrm{xc}
  \approx (\partial_\eta\Omega_\mathrm{xc}) (\partial_\mu\eta)^n$
and they thus scale at most as $\mathcal O(\mathcal N^{-4/3})$.

Using \Eq{eqn:DOmegaDMu} for $\partial_\mu\Omega_\mathrm{xc}$
and inserting the estimate
for $(\mu_\mathrm{Hxc}-\mu_\mathrm{H})$ from \Eq{eqn:MuHxcLargeN}
finally gives us an estimate of the \xc\ chemical potential
for large $\mathcal N$
\begin{multline}
  \label{eqn:FreeEnergyLargeN}
  F_\mathrm{xc}(\mathcal N) \approx \Omega_\mathrm{xc}(\mu_\mathrm{H}) \\
  + \frac{2r_\mathrm{s}}3\,(\mu_\mathrm{Hxc}-\mu_\mathrm{H})^2 \mathcal N^{\frac13}
  + \mathcal O(\mathcal N^{-\frac13}).
\end{multline}
In the thermodynamic limit the \xc\ grand potential per electron, evaluated
at the non-interacting chemical potential, is estimated to agree with the
\xc\ free energy per electron
$f_\mathrm{xc}=\lim_{\mathcal N\to\infty}F_\mathrm{xc}(\mathcal N)/\mathcal N$,
found at the interacting chemical potential.

This is the main result of this work and
the numerical studies in the following section
show that this asymptotic estimate applies already at relatively small
system sizes in the uniform electron gas for the
densities and temperatures considered.
For finite system sizes $\mathcal N$,
\Eq{eqn:FreeEnergyLargeN} relates the difference
between $\Omega_\mathrm{xc}(\mu_\mathrm{H})$ and $F_\mathrm{xc}(\mathcal N)$
to the difference between the interacting and the non-interacting chemical
potential.
The latter converges faster with system size and this
relation permits an estimate of the remaining finite-size error in
$\Omega_\mathrm{c}$ for the thermodynamic limit extrapolation.
One can also employ \Eq{eqn:MuHxcLargeN} to estimate the
interacting chemical potential $\mu_\mathrm{Hxc}$ from a correlation
calculation at a non-interacting chemical potential if the derivative with
respect to $\eta$ can be found efficiently.


\section{Numerical Results}
To assess the large system-size estimates in the previous section,
numerical calculations of the uniform electron
gas have been conducted for system sizes of
38, 54, 66, 114, 162, 246, 294, 342, 358, and 406 electrons.
The system sizes have been chosen such that
degenerate spatial orbitals can be fully occupied at zero-temperature
in a closed-shell self-consistent field calculation.

\subsection{Hartree \Sc\ field}
\label{ssc:ResultsH}
The SCF calculations in the Hartree
approximation do not include exchange, following the
scheme of RPA calculations.\cite{harl_assessing_2010}
Thus, each eigenvalue in
equation \Eq{eqn:ScfEigen} only depends on its kinetic energy and
the sum of all occupancies. A uniform shift of the eigenenergies
$\Delta\varepsilon$
is the only number that needs to be found, although in a non-linear equation.
At finite temperature, all orbitals contribute in principle.
In this work the number of spatial orbitals for the SCF calculation
has been truncated at roughly $800$ times the number of orbitals occupied
at zero temperature. Sums over the orbitals beyond this number
occurring in $\Omega_0$ and $N_\mathrm{H}$ have been approximated by integrals.
With this treatment, all SCF quantities are well converged
and the computation time for the SCF calculation is still
negligible compared to the correlation calculations.
Note that the SCF calculations have been repeated to yield relaxed
eigenenergies for each value
of the chemical potential in search for the chemical potential
$\mu_\mathrm{Hxc}$ where the expected number of
electrons matches the number of positive charges $\mathcal N$.

\begin{figure*}[t]
\begin{center}
\includegraphics{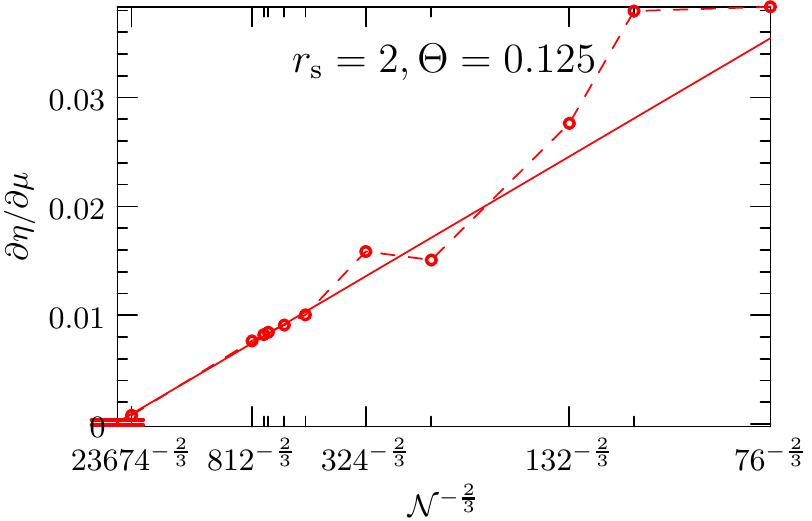}
\hfill
\includegraphics{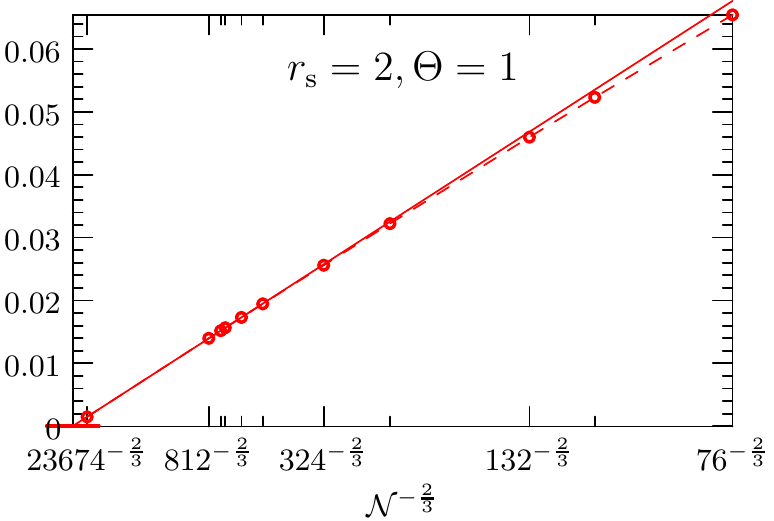}
\end{center}
\caption{%
The effective potential $\eta=\mu - \Delta\varepsilon$ is a measure for how
quantities in the perturbation expansion depend on $\mu$.
The Figure shows that the change of the effective potential $\eta$
with respect to the chemical potential $\mu$ decreases
with increasing system size. An extrapolation of the largest
calculations with $\mathcal N^{-2/3}$ indicates that the
effective potential $\eta$ becomes
independent of $\mu$ in the thermodynamic limit.
}
\label{fig:DetaDmu}
\end{figure*}

Only the difference between the eigenenergies
$\varepsilon_i=\vec k_i^2/2 + \Delta\varepsilon$
and the chemical potential $\mu$ occur in the
expressions of many-body perturbation theory
where $\Delta\varepsilon$ depends on $\mu$.
Thus, they can be viewed rather as functions of the effective chemical
potential $\eta = \mu-\Delta\varepsilon$.
Figure \ref{fig:DetaDmu} shows how the effective potential
changes when the chemical potential is changed. It plots
the derivative $\partial\eta/\partial\mu$ against
$\mathcal N^{-2/3}$ where $\mathcal N$ is the system size. The
derivative has been evaluated at the fully interacting chemical potential
$\mu_\mathrm{Hxc}$, except for the largest system size $\mathcal N=23674$,
where no correlation calculation has been conducted and $\mu_\mathrm{H}$
has been used instead.
Already for moderate system sizes, a change of the chemical potential
has about two orders of magnitude less an effect on $\eta$ and
in consequence on the expressions of FT-MBPT.
For large $\mathcal N$ the effect on $\eta$ decreases, scaling
as $\mathcal O(\mathcal N^{-2/3})$, as estimated in \Eq{eqn:deta_dmu},
and vanishes in the thermodynamic limit.

\subsection{First-order exchange}
\label{ssc:exchangeResults}
The exchange contributions to the grand potential $\Omega_\mathrm{x}$
have been evaluated according to \Eq{eqn:OmegaX} using all orbitals
that have been considered in the SCF calculation. The convergence with the number
of orbitals is faster than for the SCF quantities and no analytic treatment
of the orbitals beyond 800 times the zero-temperature orbitals is necessary.
Although considerably more demanding computationally than the SCF calculation,
its evaluation is still negligible compared to the correlation calculation.
The derivative of the exchange contribution
with respect to $\eta$ for the expected number of electrons
has been evaluated analytically.

\subsection{Linearized direct-ring coupled cluster}
\label{ssc:ldrccdResults}
The correlation and exchange effects beyond first order
have been approximated on the level of linearized direct-ring coupled
cluster doubles (ldrCCD) theory.
It is one of the simplest theories whose
zero-temperature and infinite-size limit exists. Still,
it is expected to capture the dominant part of the long range correlation.
The advantage of ldrCCD is that it can be evaluated from a
diagonalization of an effective particle/hole Hamiltonian
and consequently permits an analytic imaginary time integration.
Apart from numerical considerations, the temperature can be
arbitrarily low.

Let $N_p$ denote the number of spatial orbitals considered for
the ldrCCD calculation.
The direct-ring structure of the effective Hamiltonian in
\Eq{eqn:effectiveH} is momentum conserving.
In a uniform system and due to point-group symmetry
it is therefore sufficient to consider
independent $N_p\times N_p$ matrizes for each momentum difference
vector $\vec q=\vec k_b-\vec k_j$ in the wedge $0\leq q_x\leq q_y\leq q_z$
instead of one $N_p^2 \times N_p^2$ matrix.
For the largest system size 114 independent $4385\times 4385$ matrizes
have been diagonalized. The matrizes are real valued and symmetric
and can be diagonalized efficiently with standard linear algebra packages.

Unlike at zero-temperature, the spectrum of each matrix is not necessarily
positive-definite. Negative eigenvalues $\Lambda_F^F$ can occur
when eigenenergies of contributing
hole-orbitals are above the energies of contributing
particle-orbitals, which is possible at finite temperature.
Negative eigenvalues pose numerical difficulties occurring in
the exponent of \Eq{eqn:OmegaC}. However, the final product with
the square roots
of the occupancy products $\sqrt{n^{ab}_{ij}}$ in
Eqs.\ \eqref{eqn:EigenModeW} and \eqref{eqn:EigenModeV}
leads to a finite contribution. In practice, the term
$\delta^b_a\delta^i_j\Delta^b_j$ has been truncated to zero if the occupancy
and vacancy product $n^b_j$ was below $10^{-12}$.

For the ldrCCD calculation $N_p$ has been chosen about 20
times the number of zero-temperature
occupied spatial orbitals. The contribution from the
orbitals beyond has been extrapolated from the asymptotic behavior of
RPA-like correlation energies. The finite-basis-set error
scales as $\mathcal O(q_\mathrm{max}^3)$,
where $q_\mathrm{max}$ is the magnitude of the largest considered
plane wave momentum difference.\cite{hummel_2015}
A Hann window has been used to obtain a soft cutoff for four
different values of $q_\mathrm{max}$ to smoothen the samples
for the $q_\mathrm{max}^{-3}$ extrapolation
to the complete basis set (CBS) limit.\cite{harl_assessing_2010}
The correlation coefficients of the regression curves
range between 0.97 and practically 1.
The 67\% confidence intervals of the CBS limits are given
in the $\pm$CBS column in Table \ref{tab:FreeEnergyResults}.

\begin{figure*}[p]
\begin{center}
\includegraphics{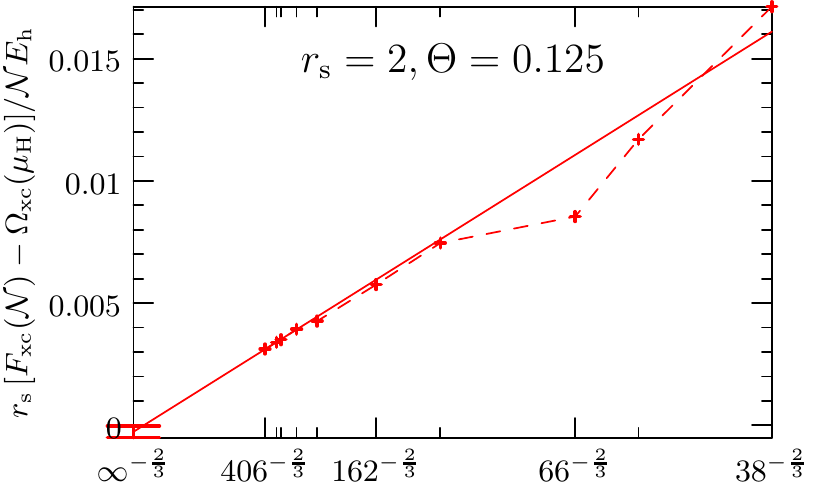} \hfill
\includegraphics{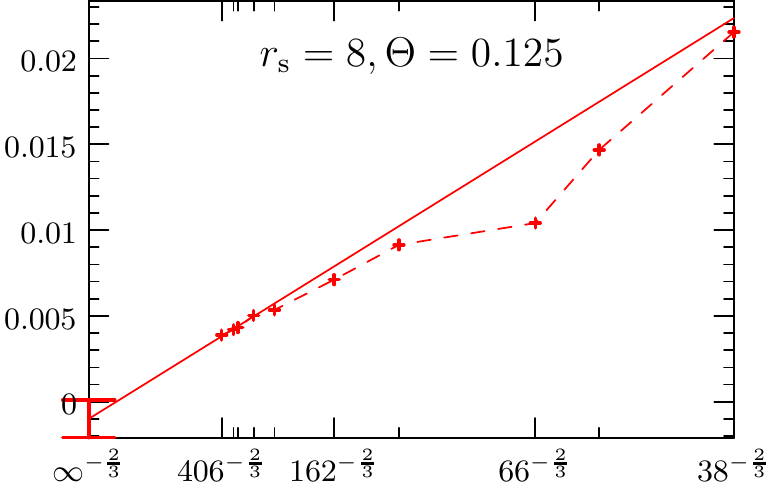} \\[1ex]
\includegraphics{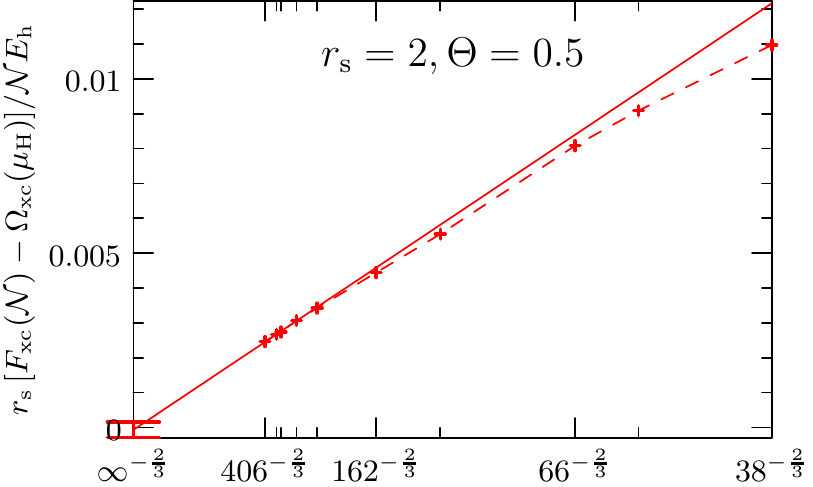} \hfill
\includegraphics{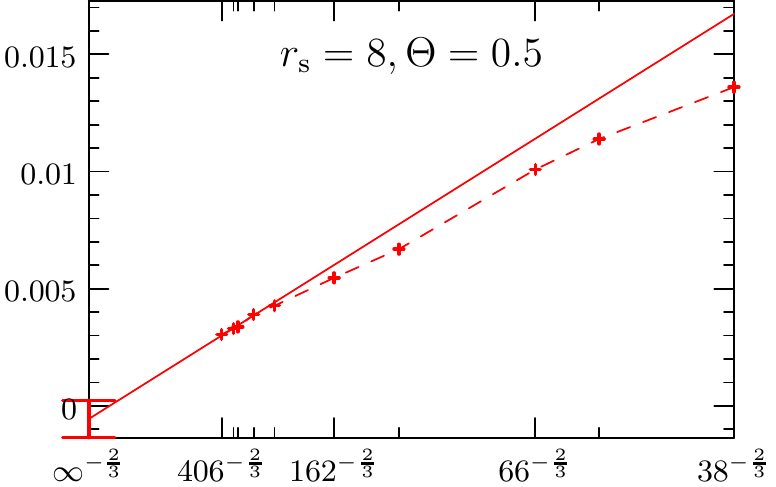} \\[1ex]
\includegraphics{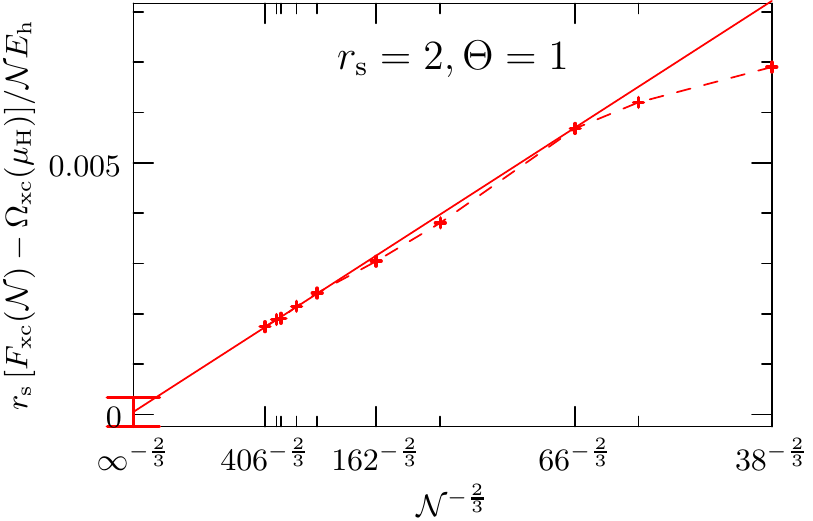} \hfill
\includegraphics{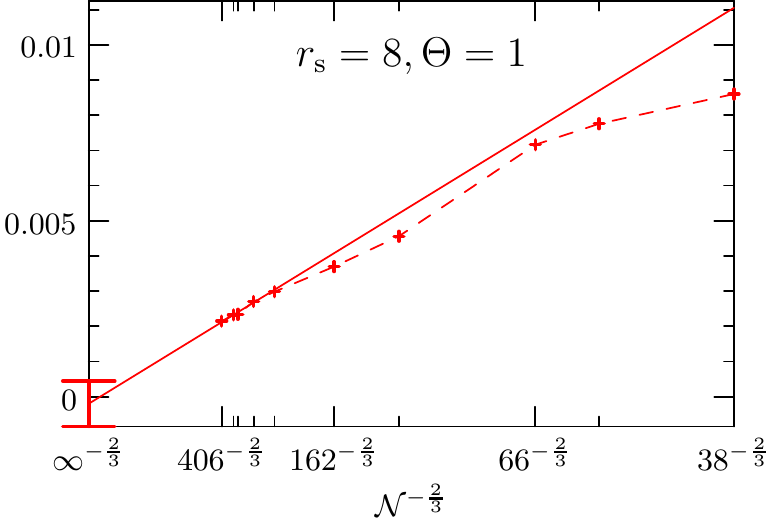}
\end{center}
\caption{%
Finite-size dependence of the
difference between the \xc\ free energy per electron
and the \xc\ grand potential per electron
for various densities and temperatures.
The correlation contributions are approximated
by the linearized direct-ring coupled cluster doubles (ldrCCD) theory.
The grand potential has been evaluated at the non-interacting chemical potential
$\mu_\mathrm{H}$ while the free energy 
requires the correlated chemical potential $\mu_\mathrm{Hxc}$.
Extrapolations of the largest system sizes show that
the two free energies coincide in the infinite-size limit.
}
\label{fig:FOmega}
\end{figure*}

Finding the thermodynamic limit poses a difficult task
in the calculation of extended systems. First, we assess
whether the asymptotic behavior estimated by
\Eq{eqn:FreeEnergyLargeN} applies in the UEG as a prototypical
warm-dense system.
For each system size, multiple calculations of $\Omega_\mathrm{Hxc}$
have been conducted in search for the chemical potential $\mu_\mathrm{Hxc}$
where $N_\mathrm{Hxc}=-\partial_\mu\Omega_\mathrm{Hxc}$ agrees with the number
of positive charges $\mathcal N$. The derivative of $\Omega_c$
has been evaluated numerically from a polynomial fit. The next
estimate $\overline{\mu}_\mathrm{Hxc}$ at the current chemical potential $\mu$
has been found from the difference of
$N_\mathrm{Hxc}-\mathcal N$
assuming that the dominant change in $N_\mathrm{Hxc}$ stems from the
change in $N_\mathrm{H}=-\partial_\mu\Omega_0$. This gives an equation
for the dominant change in the chemical potential
\begin{equation}
  \label{eqn:nextMu}
  -(\overline{\mu}_\mathrm{Hxc}- \mu)\,
    \partial_\mu\eta\,\partial^2_{\eta\mu}\Omega_0
  \approx N_\mathrm{Hxc} - \mathcal N,
\end{equation}
where all involved quantities can be readily evaluated at the
current chemical potential $\mu$.
This procedure has required about 8 iterations until convergence for
each considered system size $\mathcal N$.
Figure \ref{fig:FOmega} plots the difference between
the \xc\ free energy per electron
and the \xc\ grand potential per electron,
evaluated at the Hartree chemical potential $\mu_\mathrm{H}$,
against the system size $\mathcal N^{-2/3}$.
In this graph, an
asymptotic behavior as estimated from \Eq{eqn:FreeEnergyLargeN} is expected to
appear as a line through the origin.
As a guide to the eye,
the results are connected with dashed red lines.
The linear extrapolations from the largest
system sizes are shown as solid red lines. The 67\%-confidence
intervals of the thermodynamic limits are indicated by the error bars
on the vertical axis.
They confirm numerically that the two \xc\ free energies agree
in the thermodynamic limit of the warm UEG for all densities
and temperature considered.

\begin{figure*}[p]
\begin{center}
\includegraphics{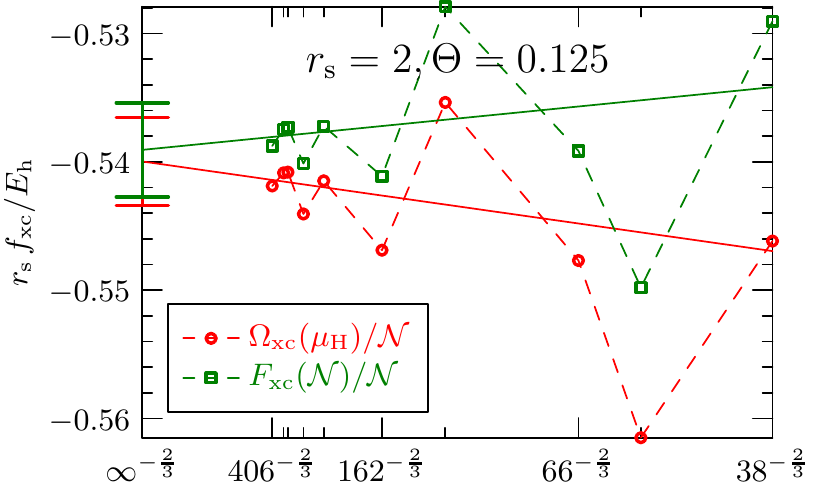} \hfill
\includegraphics{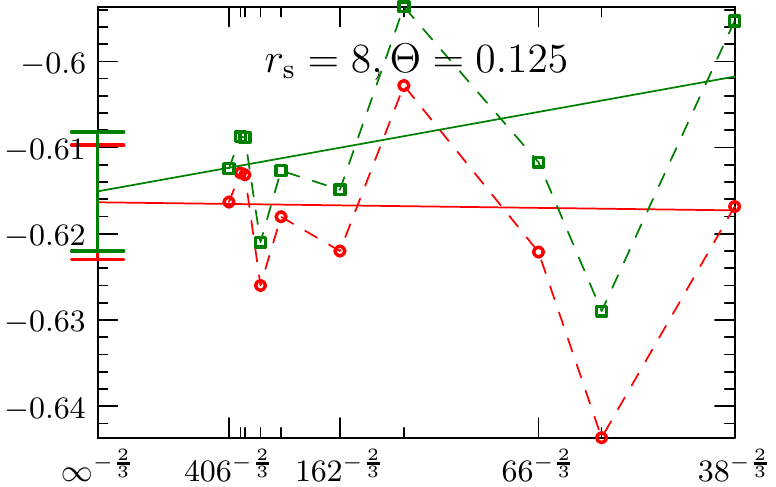} \\[1ex]
\includegraphics{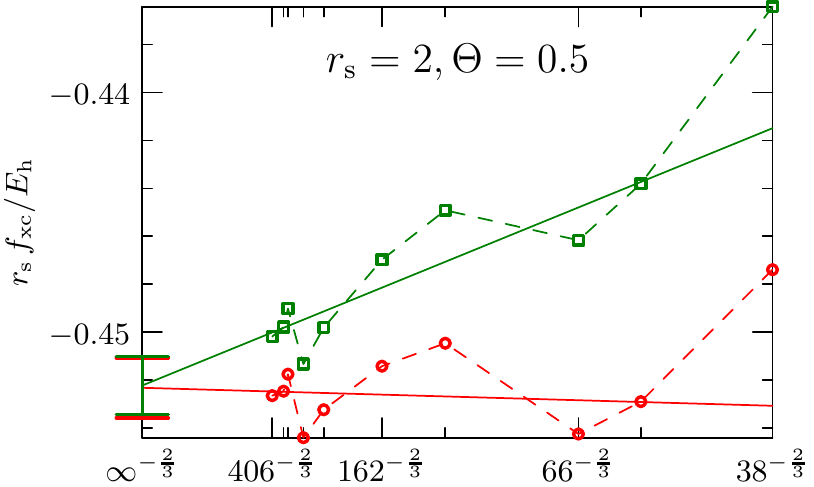} \hfill
\includegraphics{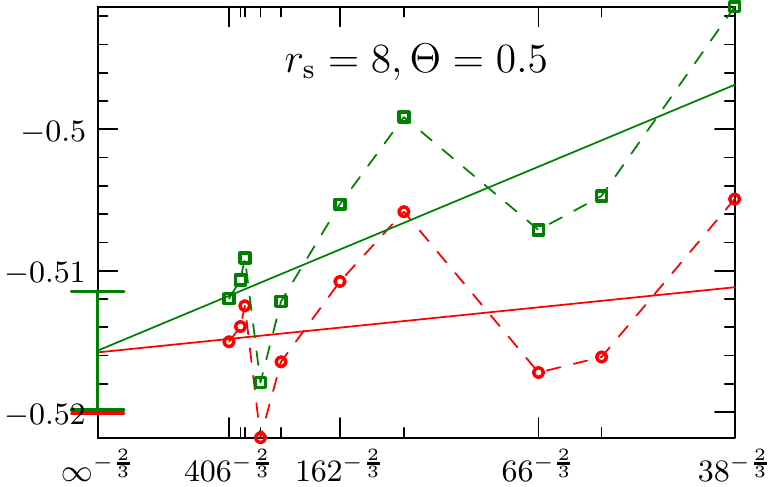} \\[1ex]
\includegraphics{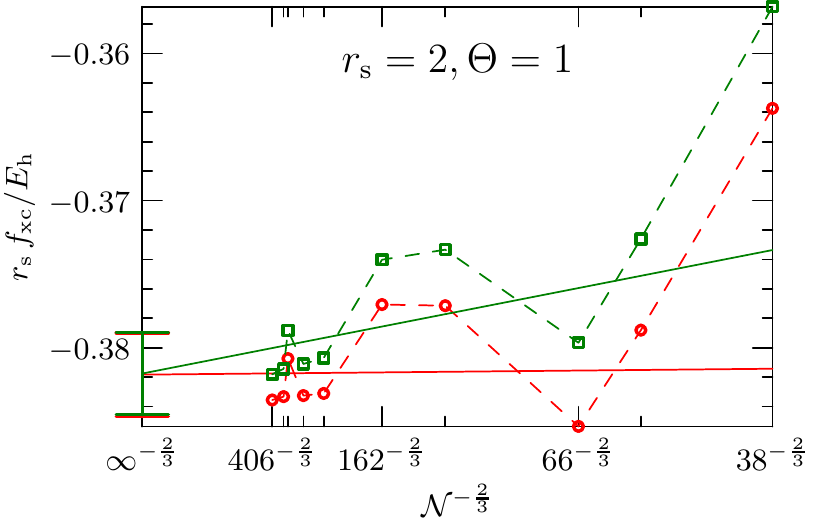} \hfill
\includegraphics{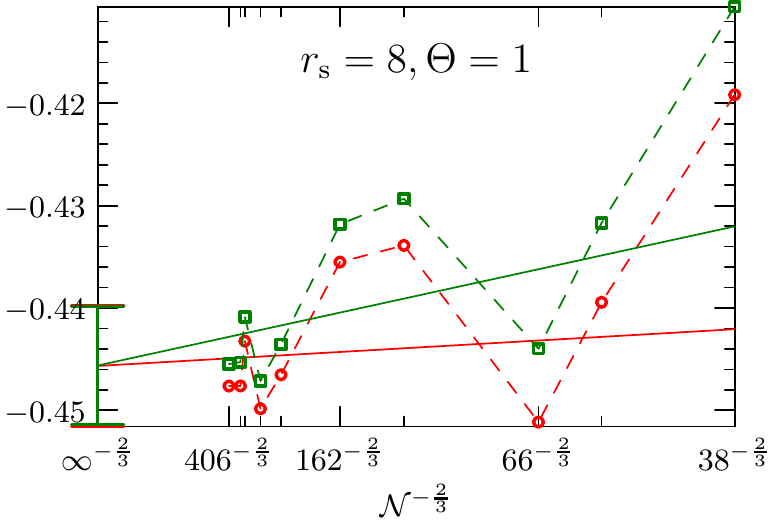}
\end{center}
\caption{%
Finite-size dependence of
the \xc\ grand potential $\Omega_\mathrm{xc}$ per electron in red
and the \xc\ free energy $F_\mathrm{xc}$ per electron in green
for various densities and temperatures.
The finite-size error of the exchange contribution is negligible for the
largest system sizes and the remaining terms are expected to converge
as $\mathcal O(\mathcal N^{-2/3})$ to the thermodynamic limit.
The solid red and green lines show the extrapolations to the
thermodynamic limit of the grand potential and the free energy, respectively.
The error bars on the vertical axes indicate the statistical error
of the extrapolations.
}
\label{fig:Fxc}
\end{figure*}

The terms in $\Omega_\mathrm{xc}$ converge with different rates
to the thermodynamic limit. At the largest considered system sizes the
exchange contributions are almost converged. The remaining correlation
terms converge as $\mathcal O(\mathcal N^{-2/3})$ in the low temperature
regime and as $\mathcal O(\mathcal N^{-1})$ otherwise.\cite{dornheim_2018}
Also, the effective chemical potential $\eta$, which $\Omega_\mathrm{c}$
depends on, converges as $\mathcal O(\mathcal N^{-2/3})$.
Thus, $F_\mathrm{xc}$ and $\Omega_\mathrm{xc}$ are also individually
expected to converge to the thermodynamic as $\mathcal O(\mathcal N^{-2/3})$.
Figure \ref{fig:Fxc} plots $F_\mathrm{xc}$ and $\Omega_\mathrm{xc}$
individually against the system size $\mathcal N^{-2/3}$.
Both contributions suffer from shell effects. They could be alleviated
by twist averaging\cite{gruber_2018,dornheim_2018} but this has not
been done in this work.
The solid lines show the $\mathcal N^{-2/3}$ fit for the largest
system sizes of the respective sets and the statistical error
of the infinite-size extrapolation for both energies is
indicated by the error bars on the vertical axes.
Interestingly, in most cases
the slope of the grand potential extrapolation is flatter
than that of the free energy extrapolation,
making the extrapolation of the grand potential less dependent
on the functional form of the asymptotic behavior.

Table \ref{tab:FreeEnergyResults} summarizes the \xc\
free energies $f_\mathrm{xc}$ found in the thermodynamic limit
and gives the 67\% confidence interval of the infinite-size extrapolation
in the $\pm$TDL column.
Despite the simple ldrCCD theory employed, the results
compare well to previous calculations, listed for instance in
Ref.\ \citenum{dornheim_2018}.
\begin{table*}[t]
\begin{center}
\begin{tabular}{|r|l|r|rrrr|}
  \hline
  \multicolumn{1}{|c|}{$r_\mathrm{s}$}
  & \multicolumn{1}{c|}{$\Theta$}
  & \multicolumn{1}{c|}{$f_\mathrm{xc}r_\mathrm{s}$}
  & \multicolumn{1}{c}{$\Omega_\mathrm{x}r_\mathrm{s}/\mathcal N$}
  & \multicolumn{1}{c}{$\Omega_\mathrm{c}r_\mathrm{s}/\mathcal N$}
  & \multicolumn{1}{c}{$\pm$TDL}
  & \multicolumn{1}{c|}{$\pm$CBS}
  \\\hline
  \multirow{3}*{2}
  & $0.125$ & $-0.5421$ & $-0.4278$ & $-0.1143$ & $\pm 0.0020$ & $\pm 0.0010$
  \\
  & $0.5$ & $-0.4528$ & $-0.2789$ & $-0.1739$ & $\pm 0.0015$ & $\pm 0.0014$
  \\
  & $1.0$ & $-0.3852$ & $-0.1739$ & $-0.2113$ & $\pm 0.0025$ & $\pm 0.0014$
  \\\hline
  \multirow{3}*{8}
  & $0.125$ & $-0.6186$ & $-0.4279$ & $-0.1907$ & $\pm 0.0050$ & $\pm 0.0041$
  \\
  & $0.5$ & $-0.5163$ & $-0.2789$ & $-0.2374$ & $\pm 0.0040$ & $\pm 0.0045$
  \\
  & $1.0$ & $-0.4522$ & $-0.1739$ & $-0.2752$ & $\pm 0.0050$ & $\pm 0.0046$\\
  \hline
\end{tabular}
\end{center}
\caption{%
Linearized direct-ring coupled cluster doubles (ldrCCD)
\xc\ free energies of the warm uniform electron
gas for various densities and temperatures. All energies are
given in Hartree.
$f_\mathrm{xc}$ is retrieved from the thermodynamic limit and
complete-basis-set limit of $\Omega_\mathrm{xc}$.
Exchange and correlation contributions have been extrapolated
separately.
The expected statistical errors from the infinite-size and infinite-basis-set
extrapolations of the correlation contributions
are given in the $\pm$TDL and $\pm$CBS column, respectively.
The TDL and CBS errors of the exchange contributions are negligible.
}
\label{tab:FreeEnergyResults}
\end{table*}


\section{Summary}
\label{ssc:Summary}
This work shows that the infinite-size limit of
finite temperature many-body perturbation theory
can be found efficiently with a truncated
Coulomb interaction.
The truncation radius is chosen
such that the volume of the interaction agrees with the volume
of the simulated cell. Such schemes have previously
been studies in classical system of electrostatically
interacting particles, as well as for the Fock-ex\-change
contribution in zero temperature MBPT.
Here, the truncation scheme is employed
for all electrostatic interactions in the uniform electron gas:
elec\-tron--elec\-tron, elec\-tron--back\-ground,
and back\-ground--back\-ground.

It is found that due to the long-ranged nature of the electrostatic interaction
the difference of the average number of mobile electrons and fixed positive
charges
scales asymptotically as $(\mu-\mu_\mathrm{Hxc})\mathcal O(\mathcal N^{1/3})$
for large system sizes
where $\mathcal N$ is the number of positive charges and
$\mu_\mathrm{Hxc}$ is the chemical
potential where the number of electrons equals the number of positive charges,
including exchange and correlation effects.
Thus, the ratio of the number of electrons and positive charges
tends to one for any finite choice of the chemical potential $\mu$.

An important consequence is that also the \xc\ grand potential
per electron, evaluated at the non-interacting
Hartree self-consistent field chemical potential
$\mu_\mathrm{H}$, asymptotically agrees with
the free energy per electron, found
from a Legendre transformation at the interacting
chemical potential $\mu_\mathrm{Hxc}$:
\[
  \frac{\Omega_\mathrm{xc}(\mu_\mathrm{H})}{\mathcal N}
  \xrightarrow{\mathcal N\to\infty}
  \frac{F_\mathrm{xc}(\mathcal N)}{\mathcal N}\,.
\]
The latter requires multiple iterations of the expensive
correlation calculations during the non-linear search for the
interacting chemical potential for each
system size considered for the thermodynamic limit extrapolation.

The above asymptotic behavior has been estimated in general
for matter under warm-dense conditions and it has been shown
explicitly for the warm uniform electron gas
for various densities and temperatures
employing the linearized direct-ring coupled cluster doubles theory
for approximating \xc\ effects.
The considered densities and temperatures
cover the region where FT-MBPT methods, such as
finite temperature coupled cluster, can complement
other methods, such as quantum Monte Carlo methods.\cite{karasiev_2019}

\section*{Data Availability}
The program and the settings used to produce the data in this
work are publicly available at
\href{https://gitlab.cc4s.org/cqt/weg-ldrccd.git}{https://gitlab.cc4s.org/cqt/weg-ldrccd.git}

\section*{Acknowledgements}
The author wishes to thank Isabella Floss, Andreas Irmler,
Evgeny Moermann,
Nikolaos Masios, Andreas Savin, Sam Trickey,
and Corbinian Wellenhofer
for constructive discussions and remarks on the manuscript.
Computer time on the computational resources of the group
of Andreas Gr\"uneis at the TU Wien
is also gratefully acknowledged.

\bibliography{mu-periodic}

\providecommand{\latin}[1]{#1}
\makeatletter
\providecommand{\doi}
  {\begingroup\let\do\@makeother\dospecials
  \catcode`\{=1 \catcode`\}=2 \doi@aux}
\providecommand{\doi@aux}[1]{\endgroup\texttt{#1}}
\makeatother
\providecommand*\mcitethebibliography{\thebibliography}
\csname @ifundefined\endcsname{endmcitethebibliography}
  {\let\endmcitethebibliography\endthebibliography}{}
\begin{mcitethebibliography}{55}
\providecommand*\natexlab[1]{#1}
\providecommand*\mciteSetBstSublistMode[1]{}
\providecommand*\mciteSetBstMaxWidthForm[2]{}
\providecommand*\mciteBstWouldAddEndPuncttrue
  {\def\EndOfBibitem{\unskip.}}
\providecommand*\mciteBstWouldAddEndPunctfalse
  {\let\EndOfBibitem\relax}
\providecommand*\mciteSetBstMidEndSepPunct[3]{}
\providecommand*\mciteSetBstSublistLabelBeginEnd[3]{}
\providecommand*\EndOfBibitem{}
\mciteSetBstSublistMode{f}
\mciteSetBstMaxWidthForm{subitem}{(\alph{mcitesubitemcount})}
\mciteSetBstSublistLabelBeginEnd
  {\mcitemaxwidthsubitemform\space}
  {\relax}
  {\relax}

\bibitem[Graziani \latin{et~al.}(2014)Graziani, Desjarlais, Redmer, and
  Trickey]{graziani_2014}
Graziani,~F., Desjarlais,~M.~P., Redmer,~R., Trickey,~S.~B., Eds.
  \emph{Frontiers and Challenges in Warm Dense Matter}; Springer International
  Publishing, 2014\relax
\mciteBstWouldAddEndPuncttrue
\mciteSetBstMidEndSepPunct{\mcitedefaultmidpunct}
{\mcitedefaultendpunct}{\mcitedefaultseppunct}\relax
\EndOfBibitem
\bibitem[Iyer \latin{et~al.}(2015)Iyer, Srednicki, and Rigol]{iyer_2015}
Iyer,~D.; Srednicki,~M.; Rigol,~M. Optimization of finite-size errors in
  finite-temperature calculations of unordered phases. \emph{Phys. Rev. E}
  \textbf{2015}, \emph{91}, 062142\relax
\mciteBstWouldAddEndPuncttrue
\mciteSetBstMidEndSepPunct{\mcitedefaultmidpunct}
{\mcitedefaultendpunct}{\mcitedefaultseppunct}\relax
\EndOfBibitem
\bibitem[Brown \latin{et~al.}(2013)Brown, Clark, DuBois, and
  Ceperley]{brown_2013}
Brown,~E.~W.; Clark,~B.~K.; DuBois,~J.~L.; Ceperley,~D.~M. Path-Integral Monte
  Carlo Simulation of the Warm Dense Homogeneous Electron Gas. \emph{Phys. Rev.
  Lett.} \textbf{2013}, \emph{110}, 146405\relax
\mciteBstWouldAddEndPuncttrue
\mciteSetBstMidEndSepPunct{\mcitedefaultmidpunct}
{\mcitedefaultendpunct}{\mcitedefaultseppunct}\relax
\EndOfBibitem
\bibitem[Militzer \latin{et~al.}(2019)Militzer, Pollock, and
  Ceperley]{militzer_2019}
Militzer,~B.; Pollock,~E.; Ceperley,~D. Path integral Monte Carlo calculation
  of the momentum distribution of the homogeneous electron gas at finite
  temperature. \emph{High Energy Density Physics} \textbf{2019}, \emph{30}, 13
  -- 20\relax
\mciteBstWouldAddEndPuncttrue
\mciteSetBstMidEndSepPunct{\mcitedefaultmidpunct}
{\mcitedefaultendpunct}{\mcitedefaultseppunct}\relax
\EndOfBibitem
\bibitem[Fetter and Walecka(2003)Fetter, and Walecka]{fetter_quantum_2003}
Fetter,~A.~L.; Walecka,~J.~D. \emph{Quantum theory of many-particle systems};
  Dover Publications: Mineola, N.Y, 2003\relax
\mciteBstWouldAddEndPuncttrue
\mciteSetBstMidEndSepPunct{\mcitedefaultmidpunct}
{\mcitedefaultendpunct}{\mcitedefaultseppunct}\relax
\EndOfBibitem
\bibitem[Thouless(2014)]{thouless_2014}
Thouless,~D.~J. \emph{The quantum mechanics of many-body systems}, second dover
  edition ed.; Dover Publications, Inc, 2014\relax
\mciteBstWouldAddEndPuncttrue
\mciteSetBstMidEndSepPunct{\mcitedefaultmidpunct}
{\mcitedefaultendpunct}{\mcitedefaultseppunct}\relax
\EndOfBibitem
\bibitem[Kohn and Luttinger(1960)Kohn, and Luttinger]{kohn_ground-state_1960}
Kohn,~W.; Luttinger,~J.~M. Ground-{State} {Energy} of a {Many}-{Fermion}
  {System}. \emph{Phys. Rev.} \textbf{1960}, \emph{118}, 41--45\relax
\mciteBstWouldAddEndPuncttrue
\mciteSetBstMidEndSepPunct{\mcitedefaultmidpunct}
{\mcitedefaultendpunct}{\mcitedefaultseppunct}\relax
\EndOfBibitem
\bibitem[Liang \latin{et~al.}(2015)Liang, Xu, and Xing]{Liang_2015}
Liang,~Y.; Xu,~Z.; Xing,~X. A multi-scale Monte Carlo method for electrolytes.
  \emph{New Journal of Physics} \textbf{2015}, \emph{17}\relax
\mciteBstWouldAddEndPuncttrue
\mciteSetBstMidEndSepPunct{\mcitedefaultmidpunct}
{\mcitedefaultendpunct}{\mcitedefaultseppunct}\relax
\EndOfBibitem
\bibitem[Spencer and Alavi(2008)Spencer, and Alavi]{spencer_2008}
Spencer,~J.; Alavi,~A. Efficient calculation of the exact exchange energy in
  periodic systems using a truncated Coulomb potential. \emph{Phys. Rev. B}
  \textbf{2008}, \emph{77}, 193110\relax
\mciteBstWouldAddEndPuncttrue
\mciteSetBstMidEndSepPunct{\mcitedefaultmidpunct}
{\mcitedefaultendpunct}{\mcitedefaultseppunct}\relax
\EndOfBibitem
\bibitem[Gygi and Baldereschi(1986)Gygi, and Baldereschi]{gygi_1986}
Gygi,~F.; Baldereschi,~A. Self-consistent Hartree-Fock and screened-exchange
  calculations in solids: Application to silicon. \emph{Phys. Rev. B}
  \textbf{1986}, \emph{34}, 4405--4408\relax
\mciteBstWouldAddEndPuncttrue
\mciteSetBstMidEndSepPunct{\mcitedefaultmidpunct}
{\mcitedefaultendpunct}{\mcitedefaultseppunct}\relax
\EndOfBibitem
\bibitem[Carrier \latin{et~al.}(2007)Carrier, Rohra, and
  G\"orling]{carrier_2007}
Carrier,~P.; Rohra,~S.; G\"orling,~A. General treatment of the singularities in
  Hartree-Fock and exact-exchange Kohn-Sham methods for solids. \emph{Phys.
  Rev. B} \textbf{2007}, \emph{75}, 205126\relax
\mciteBstWouldAddEndPuncttrue
\mciteSetBstMidEndSepPunct{\mcitedefaultmidpunct}
{\mcitedefaultendpunct}{\mcitedefaultseppunct}\relax
\EndOfBibitem
\bibitem[Irmler \latin{et~al.}(2018)Irmler, Burow, and Pauly]{irmler_2018}
Irmler,~A.; Burow,~A.~M.; Pauly,~F. Robust Periodic Fock Exchange with
  Atom-Centered Gaussian Basis Sets. \emph{Journal of Chemical Theory and
  Computation} \textbf{2018}, \emph{14}, 4567--4580, PMID: 30080979\relax
\mciteBstWouldAddEndPuncttrue
\mciteSetBstMidEndSepPunct{\mcitedefaultmidpunct}
{\mcitedefaultendpunct}{\mcitedefaultseppunct}\relax
\EndOfBibitem
\bibitem[Sundararaman and Arias(2013)Sundararaman, and
  Arias]{sundararaman_2013}
Sundararaman,~R.; Arias,~T.~A. Regularization of the Coulomb singularity in
  exact exchange by Wigner-Seitz truncated interactions: Towards chemical
  accuracy in nontrivial systems. \emph{Phys. Rev. B} \textbf{2013}, \emph{87},
  165122\relax
\mciteBstWouldAddEndPuncttrue
\mciteSetBstMidEndSepPunct{\mcitedefaultmidpunct}
{\mcitedefaultendpunct}{\mcitedefaultseppunct}\relax
\EndOfBibitem
\bibitem[Matsubara(1955)]{matsubara_new_1955}
Matsubara,~T. A {New} {Approach} to {Quantum}-{Statistical} {Mechanics}.
  \emph{Prog. Theor. Phys.} \textbf{1955}, \emph{14}, 351--378\relax
\mciteBstWouldAddEndPuncttrue
\mciteSetBstMidEndSepPunct{\mcitedefaultmidpunct}
{\mcitedefaultendpunct}{\mcitedefaultseppunct}\relax
\EndOfBibitem
\bibitem[Bloch and De~Dominicis(1958)Bloch, and De~Dominicis]{bloch_1958}
Bloch,~C.; De~Dominicis,~C. Un d\'{e}veloppement du potentiel de gibbs d'un
  syst\`{e}me quantique compos\'{e} d'un grand nombre de particules.
  \emph{Nuclear Physics} \textbf{1958}, \emph{7}, 459--479\relax
\mciteBstWouldAddEndPuncttrue
\mciteSetBstMidEndSepPunct{\mcitedefaultmidpunct}
{\mcitedefaultendpunct}{\mcitedefaultseppunct}\relax
\EndOfBibitem
\bibitem[Bloch and De~Dominicis(1959)Bloch, and De~Dominicis]{bloch_1959}
Bloch,~C.; De~Dominicis,~C. Un d\'{e}veloppement du potentiel de Gibbs d'un
  syst\`{e}me compos\'{e} d'un grand nombre de particules (II). \emph{Nuclear
  Physics} \textbf{1959}, \emph{10}, 181--196\relax
\mciteBstWouldAddEndPuncttrue
\mciteSetBstMidEndSepPunct{\mcitedefaultmidpunct}
{\mcitedefaultendpunct}{\mcitedefaultseppunct}\relax
\EndOfBibitem
\bibitem[Nettelmann \latin{et~al.}(2008)Nettelmann, Redmer, and
  Blaschke]{nettelmann_2008}
Nettelmann,~N.; Redmer,~R.; Blaschke,~D. Warm dense matter in giant planets and
  exoplanets. \emph{Physics of Particles and Nuclei} \textbf{2008}, \emph{39},
  1122--1127\relax
\mciteBstWouldAddEndPuncttrue
\mciteSetBstMidEndSepPunct{\mcitedefaultmidpunct}
{\mcitedefaultendpunct}{\mcitedefaultseppunct}\relax
\EndOfBibitem
\bibitem[Hirata and He(2013)Hirata, and He]{hirata_kohnluttinger_2013}
Hirata,~S.; He,~X. On the {Kohn}–{Luttinger} conundrum. \emph{J. Chem. Phys.}
  \textbf{2013}, \emph{138}, 204112\relax
\mciteBstWouldAddEndPuncttrue
\mciteSetBstMidEndSepPunct{\mcitedefaultmidpunct}
{\mcitedefaultendpunct}{\mcitedefaultseppunct}\relax
\EndOfBibitem
\bibitem[Son \latin{et~al.}(2014)Son, Thiele, Jurek, Ziaja, and
  Santra]{son_2014}
Son,~S.-K.; Thiele,~R.; Jurek,~Z.; Ziaja,~B.; Santra,~R. Quantum-Mechanical
  Calculation of Ionization-Potential Lowering in Dense Plasmas. \emph{Phys.
  Rev. X} \textbf{2014}, \emph{4}, 031004\relax
\mciteBstWouldAddEndPuncttrue
\mciteSetBstMidEndSepPunct{\mcitedefaultmidpunct}
{\mcitedefaultendpunct}{\mcitedefaultseppunct}\relax
\EndOfBibitem
\bibitem[Santra and Schirmer(2017)Santra, and Schirmer]{santra_2017}
Santra,~R.; Schirmer,~J. Finite-temperature second-order many-body perturbation
  theory revisited. \emph{Chem. Phys.} \textbf{2017}, \emph{482},
  355--361\relax
\mciteBstWouldAddEndPuncttrue
\mciteSetBstMidEndSepPunct{\mcitedefaultmidpunct}
{\mcitedefaultendpunct}{\mcitedefaultseppunct}\relax
\EndOfBibitem
\bibitem[Gupta and Rajagopal(1980)Gupta, and Rajagopal]{gupta_1980}
Gupta,~U.; Rajagopal,~A.~K. Exchange-correlation potential for inhomogeneous
  electron systems at finite temperatures. \emph{Phys. Rev. A} \textbf{1980},
  \emph{22}, 2792--2797\relax
\mciteBstWouldAddEndPuncttrue
\mciteSetBstMidEndSepPunct{\mcitedefaultmidpunct}
{\mcitedefaultendpunct}{\mcitedefaultseppunct}\relax
\EndOfBibitem
\bibitem[Perrot(1982)]{perrot_1982}
Perrot,~F. Temperature-dependent nonlinear screening of a proton in an electron
  gas. \emph{Phys. Rev. A} \textbf{1982}, \emph{25}, 489--495\relax
\mciteBstWouldAddEndPuncttrue
\mciteSetBstMidEndSepPunct{\mcitedefaultmidpunct}
{\mcitedefaultendpunct}{\mcitedefaultseppunct}\relax
\EndOfBibitem
\bibitem[Perrot and Dharma-wardana(1984)Perrot, and
  Dharma-wardana]{perrot_1984}
Perrot,~F.; Dharma-wardana,~M. W.~C. Exchange and correlation potentials for
  electron-ion systems at finite temperatures. \emph{Phys. Rev. A}
  \textbf{1984}, \emph{30}, 2619--2626\relax
\mciteBstWouldAddEndPuncttrue
\mciteSetBstMidEndSepPunct{\mcitedefaultmidpunct}
{\mcitedefaultendpunct}{\mcitedefaultseppunct}\relax
\EndOfBibitem
\bibitem[Csanak and Kilcrease(1997)Csanak, and Kilcrease]{csanak_1997}
Csanak,~G.; Kilcrease,~D. Photoabsorption in hot, dense plasmas\textemdash{}The
  average atom, the spherical cell model, and the random phase approximation.
  \emph{J. Quant. Spectrosc. Radiat. Transf.} \textbf{1997}, \emph{58},
  537--551\relax
\mciteBstWouldAddEndPuncttrue
\mciteSetBstMidEndSepPunct{\mcitedefaultmidpunct}
{\mcitedefaultendpunct}{\mcitedefaultseppunct}\relax
\EndOfBibitem
\bibitem[van Leeuwen \latin{et~al.}(2006)van Leeuwen, Dahlen, and
  Stan]{vanLeeuwen_2006}
van Leeuwen,~R.; Dahlen,~N.~E.; Stan,~A. Total energies from variational
  functionals of the Green function and the renormalized four-point vertex.
  \emph{Phys. Rev. B} \textbf{2006}, \emph{74}, 195105\relax
\mciteBstWouldAddEndPuncttrue
\mciteSetBstMidEndSepPunct{\mcitedefaultmidpunct}
{\mcitedefaultendpunct}{\mcitedefaultseppunct}\relax
\EndOfBibitem
\bibitem[Welden \latin{et~al.}(2016)Welden, Rusakov, and Zgid]{welden_2016}
Welden,~A.~R.; Rusakov,~A.~A.; Zgid,~D. Exploring connections between
  statistical mechanics and Green's functions for realistic systems:
  Temperature dependent electronic entropy and internal energy from a
  self-consistent second-order Green's function. \emph{J. Chem. Phys.}
  \textbf{2016}, \emph{145}\relax
\mciteBstWouldAddEndPuncttrue
\mciteSetBstMidEndSepPunct{\mcitedefaultmidpunct}
{\mcitedefaultendpunct}{\mcitedefaultseppunct}\relax
\EndOfBibitem
\bibitem[Mandal \latin{et~al.}(2003)Mandal, Ghosh, Sanyal, and
  Mukherjee]{mandal_2003}
Mandal,~S.~H.; Ghosh,~R.; Sanyal,~G.; Mukherjee,~D. A finite-temperature
  generalisation of the coupled cluster method: a non-perturbative access to
  grand partition functions. \emph{Int. J. Mod. Phys. B} \textbf{2003},
  \emph{17}, 5367--5377\relax
\mciteBstWouldAddEndPuncttrue
\mciteSetBstMidEndSepPunct{\mcitedefaultmidpunct}
{\mcitedefaultendpunct}{\mcitedefaultseppunct}\relax
\EndOfBibitem
\bibitem[White and Chan(2018)White, and Chan]{white_2018}
White,~A.~F.; Chan,~G. K.-L. A Time-Dependent Formulation of Coupled-Cluster
  Theory for Many-Fermion Systems at Finite Temperature. \emph{Journal of
  Chemical Theory and Computation} \textbf{2018}, \emph{14}, 5690--5700\relax
\mciteBstWouldAddEndPuncttrue
\mciteSetBstMidEndSepPunct{\mcitedefaultmidpunct}
{\mcitedefaultendpunct}{\mcitedefaultseppunct}\relax
\EndOfBibitem
\bibitem[White and Kin-Lic~Chan(2020)White, and Kin-Lic~Chan]{white_2020}
White,~A.~F.; Kin-Lic~Chan,~G. Finite-temperature coupled cluster: Efficient
  implementation and application to prototypical systems. \emph{The Journal of
  Chemical Physics} \textbf{2020}, \emph{152}, 224104\relax
\mciteBstWouldAddEndPuncttrue
\mciteSetBstMidEndSepPunct{\mcitedefaultmidpunct}
{\mcitedefaultendpunct}{\mcitedefaultseppunct}\relax
\EndOfBibitem
\bibitem[Hummel(2018)]{hummel_2018}
Hummel,~F. Finite Temperature Coupled Cluster Theories for Extended Systems.
  \emph{Journal of Chemical Theory and Computation} \textbf{2018}, \emph{14},
  6505--6514\relax
\mciteBstWouldAddEndPuncttrue
\mciteSetBstMidEndSepPunct{\mcitedefaultmidpunct}
{\mcitedefaultendpunct}{\mcitedefaultseppunct}\relax
\EndOfBibitem
\bibitem[Harsha \latin{et~al.}(2019)Harsha, Henderson, and
  Scuseria]{harsha_2019a}
Harsha,~G.; Henderson,~T.~M.; Scuseria,~G.~E. Thermofield theory for
  finite-temperature quantum chemistry. \emph{The Journal of Chemical Physics}
  \textbf{2019}, \emph{150}\relax
\mciteBstWouldAddEndPuncttrue
\mciteSetBstMidEndSepPunct{\mcitedefaultmidpunct}
{\mcitedefaultendpunct}{\mcitedefaultseppunct}\relax
\EndOfBibitem
\bibitem[Harsha \latin{et~al.}(2019)Harsha, Henderson, and
  Scuseria]{harsha_2019b}
Harsha,~G.; Henderson,~T.~M.; Scuseria,~G.~E. Thermofield Theory for
  Finite-Temperature Coupled Cluster. \emph{Journal of Chemical Theory and
  Computation} \textbf{2019}, \emph{15}, 6127--6136\relax
\mciteBstWouldAddEndPuncttrue
\mciteSetBstMidEndSepPunct{\mcitedefaultmidpunct}
{\mcitedefaultendpunct}{\mcitedefaultseppunct}\relax
\EndOfBibitem
\bibitem[Harsha \latin{et~al.}(2022)Harsha, Xu, Henderson, and
  Scuseria]{harsha_2022}
Harsha,~G.; Xu,~Y.; Henderson,~T.~M.; Scuseria,~G.~E. Thermal coupled cluster
  theory for SU(2) systems. \emph{Phys. Rev. B} \textbf{2022}, \emph{105},
  045125\relax
\mciteBstWouldAddEndPuncttrue
\mciteSetBstMidEndSepPunct{\mcitedefaultmidpunct}
{\mcitedefaultendpunct}{\mcitedefaultseppunct}\relax
\EndOfBibitem
\bibitem[Hirata and Jha(2019)Hirata, and Jha]{hirata_2019b}
Hirata,~S.; Jha,~P.~K. In \emph{Chapter Two - Converging finite-temperature
  many-body perturbation theory in the grand canonical ensemble that conserves
  the average number of electrons}; Dixon,~D.~A., Ed.; Annual Reports in
  Computational Chemistry; Elsevier, 2019; Vol.~15; pp 17 -- 37\relax
\mciteBstWouldAddEndPuncttrue
\mciteSetBstMidEndSepPunct{\mcitedefaultmidpunct}
{\mcitedefaultendpunct}{\mcitedefaultseppunct}\relax
\EndOfBibitem
\bibitem[Jha and Hirata(2020)Jha, and Hirata]{hirata_2020}
Jha,~P.~K.; Hirata,~S. Finite-temperature many-body perturbation theory in the
  canonical ensemble. \emph{Physical Review E} \textbf{2020}, \emph{101}\relax
\mciteBstWouldAddEndPuncttrue
\mciteSetBstMidEndSepPunct{\mcitedefaultmidpunct}
{\mcitedefaultendpunct}{\mcitedefaultseppunct}\relax
\EndOfBibitem
\bibitem[Harsha \latin{et~al.}(2020)Harsha, Henderson, and
  Scuseria]{harsha_2020}
Harsha,~G.; Henderson,~T.~M.; Scuseria,~G.~E. Wave function methods for
  canonical ensemble thermal averages in correlated many-fermion systems.
  \emph{The Journal of Chemical Physics} \textbf{2020}, \emph{153},
  124115\relax
\mciteBstWouldAddEndPuncttrue
\mciteSetBstMidEndSepPunct{\mcitedefaultmidpunct}
{\mcitedefaultendpunct}{\mcitedefaultseppunct}\relax
\EndOfBibitem
\bibitem[Mermin(1963)]{mermin_1963}
Mermin,~N. Stability of the thermal Hartree-Fock approximation. \emph{Ann.
  Phys. (N.~Y.)} \textbf{1963}, \emph{21}, 99 -- 121\relax
\mciteBstWouldAddEndPuncttrue
\mciteSetBstMidEndSepPunct{\mcitedefaultmidpunct}
{\mcitedefaultendpunct}{\mcitedefaultseppunct}\relax
\EndOfBibitem
\bibitem[Mermin(1965)]{mermin_1965}
Mermin,~N. Thermal Properties of the Inhomogeneous Electron Gas. \emph{Phys.
  Rev. A} \textbf{1965}, \emph{137}, 1441\relax
\mciteBstWouldAddEndPuncttrue
\mciteSetBstMidEndSepPunct{\mcitedefaultmidpunct}
{\mcitedefaultendpunct}{\mcitedefaultseppunct}\relax
\EndOfBibitem
\bibitem[Pittalis \latin{et~al.}(2011)Pittalis, Proetto, Floris, Sanna,
  Bersier, Burke, and Gross]{pittalis_2011}
Pittalis,~S.; Proetto,~C.~R.; Floris,~A.; Sanna,~A.; Bersier,~C.; Burke,~K.;
  Gross,~E. K.~U. Exact Conditions in Finite-Temperature Density-Functional
  Theory. \emph{Phys. Rev. Lett.} \textbf{2011}, \emph{107}, 163001\relax
\mciteBstWouldAddEndPuncttrue
\mciteSetBstMidEndSepPunct{\mcitedefaultmidpunct}
{\mcitedefaultendpunct}{\mcitedefaultseppunct}\relax
\EndOfBibitem
\bibitem[Karasiev \latin{et~al.}(2016)Karasiev, Calder\'{\i}n, and
  Trickey]{karasiev_2016}
Karasiev,~V.~V.; Calder\'{\i}n,~L.; Trickey,~S.~B. Importance of
  finite-temperature exchange correlation for warm dense matter calculations.
  \emph{Phys. Rev. E} \textbf{2016}, \emph{93}, 063207\relax
\mciteBstWouldAddEndPuncttrue
\mciteSetBstMidEndSepPunct{\mcitedefaultmidpunct}
{\mcitedefaultendpunct}{\mcitedefaultseppunct}\relax
\EndOfBibitem
\bibitem[Karasiev \latin{et~al.}(2014)Karasiev, Sjostrom, Chakraborty, Dufty,
  Runge, Harris, and Trickey]{karasiev_2014}
Karasiev,~V.~V.; Sjostrom,~T.; Chakraborty,~D.; Dufty,~J.~W.; Runge,~K.;
  Harris,~F.~E.; Trickey,~S.~B. In \emph{Frontiers and Challenges in Warm Dense
  Matter}; Graziani,~F., Desjarlais,~M.~P., Redmer,~R., Trickey,~S.~B., Eds.;
  Springer International Publishing, 2014\relax
\mciteBstWouldAddEndPuncttrue
\mciteSetBstMidEndSepPunct{\mcitedefaultmidpunct}
{\mcitedefaultendpunct}{\mcitedefaultseppunct}\relax
\EndOfBibitem
\bibitem[Luo \latin{et~al.}(2020)Luo, Karasiev, and Trickey]{luo_2020}
Luo,~K.; Karasiev,~V.~V.; Trickey,~S.~B. Towards accurate orbital-free
  simulations: A generalized gradient approximation for the noninteracting free
  energy density functional. \emph{Phys. Rev. B} \textbf{2020}, \emph{101},
  075116\relax
\mciteBstWouldAddEndPuncttrue
\mciteSetBstMidEndSepPunct{\mcitedefaultmidpunct}
{\mcitedefaultendpunct}{\mcitedefaultseppunct}\relax
\EndOfBibitem
\bibitem[Jha and Hirata(2019)Jha, and Hirata]{hirata_2019a}
Jha,~P.~K.; Hirata,~S. In \emph{Chapter One - Numerical evidence invalidating
  finite-temperature many-body perturbation theory}; Dixon,~D.~A., Ed.; Annual
  Reports in Computational Chemistry; Elsevier, 2019; Vol.~15; pp 3 -- 15\relax
\mciteBstWouldAddEndPuncttrue
\mciteSetBstMidEndSepPunct{\mcitedefaultmidpunct}
{\mcitedefaultendpunct}{\mcitedefaultseppunct}\relax
\EndOfBibitem
\bibitem[Sjostrom and Dufty(2013)Sjostrom, and Dufty]{Sjostrom_2013}
Sjostrom,~T.; Dufty,~J. Uniform electron gas at finite temperatures.
  \emph{Physical Review B} \textbf{2013}, \emph{88}\relax
\mciteBstWouldAddEndPuncttrue
\mciteSetBstMidEndSepPunct{\mcitedefaultmidpunct}
{\mcitedefaultendpunct}{\mcitedefaultseppunct}\relax
\EndOfBibitem
\bibitem[Dornheim \latin{et~al.}(2018)Dornheim, Groth, and
  Bonitz]{dornheim_2018}
Dornheim,~T.; Groth,~S.; Bonitz,~M. The uniform electron gas at warm dense
  matter conditions. \emph{Physics Reports} \textbf{2018}, \emph{744}, 1 --
  86\relax
\mciteBstWouldAddEndPuncttrue
\mciteSetBstMidEndSepPunct{\mcitedefaultmidpunct}
{\mcitedefaultendpunct}{\mcitedefaultseppunct}\relax
\EndOfBibitem
\bibitem[Karasiev \latin{et~al.}(2019)Karasiev, Trickey, and
  Dufty]{karasiev_2019}
Karasiev,~V.~V.; Trickey,~S.~B.; Dufty,~J.~W. Status of free-energy
  representations for the homogeneous electron gas. \emph{Physical Review B}
  \textbf{2019}, \emph{99}\relax
\mciteBstWouldAddEndPuncttrue
\mciteSetBstMidEndSepPunct{\mcitedefaultmidpunct}
{\mcitedefaultendpunct}{\mcitedefaultseppunct}\relax
\EndOfBibitem
\bibitem[Wellenhofer(2019)]{wellenhofer_2019}
Wellenhofer,~C. Zero-temperature limit and statistical quasiparticles in
  many-body perturbation theory. \emph{Physical Review C} \textbf{2019},
  \emph{99}\relax
\mciteBstWouldAddEndPuncttrue
\mciteSetBstMidEndSepPunct{\mcitedefaultmidpunct}
{\mcitedefaultendpunct}{\mcitedefaultseppunct}\relax
\EndOfBibitem
\bibitem[Hirata(2022)]{hirata_2022}
Hirata,~S. General solution to the Kohn–Luttinger nonconvergence problem.
  \emph{Chemical Physics Letters} \textbf{2022}, \emph{800}, 139668\relax
\mciteBstWouldAddEndPuncttrue
\mciteSetBstMidEndSepPunct{\mcitedefaultmidpunct}
{\mcitedefaultendpunct}{\mcitedefaultseppunct}\relax
\EndOfBibitem
\bibitem[Harl \latin{et~al.}(2010)Harl, Schimka, and
  Kresse]{harl_assessing_2010}
Harl,~J.; Schimka,~L.; Kresse,~G. Assessing the quality of the random phase
  approximation for lattice constants and atomization energies of solids.
  \emph{Physical Review B} \textbf{2010}, \emph{81}\relax
\mciteBstWouldAddEndPuncttrue
\mciteSetBstMidEndSepPunct{\mcitedefaultmidpunct}
{\mcitedefaultendpunct}{\mcitedefaultseppunct}\relax
\EndOfBibitem
\bibitem[Mattuck(1992)]{mattuck_1992}
Mattuck,~R.~D. \emph{A Guide to Feynman Diagrams in the Many-Body Problem};
  Dover Publications: Mineola, N.Y, 1992\relax
\mciteBstWouldAddEndPuncttrue
\mciteSetBstMidEndSepPunct{\mcitedefaultmidpunct}
{\mcitedefaultendpunct}{\mcitedefaultseppunct}\relax
\EndOfBibitem
\bibitem[Macke(1950)]{macke_uber_1950}
Macke,~W. Über die {Wechselwirkungen} im {Fermi}-{Gas},
  {Polarisationserscheinungen}, {Correlationsenergie},
  {Elektronenkondensation}. \emph{Z. Naturforsch.} \textbf{1950}, \emph{5a},
  192--208\relax
\mciteBstWouldAddEndPuncttrue
\mciteSetBstMidEndSepPunct{\mcitedefaultmidpunct}
{\mcitedefaultendpunct}{\mcitedefaultseppunct}\relax
\EndOfBibitem
\bibitem[Pines and Bohm(1952)Pines, and Bohm]{pines_collective_1952}
Pines,~D.; Bohm,~D. A {Collective} {Description} of {Electron} {Interactions}:
  {II}. {Collective} vs {Individual} {Particle} {Aspects} of the
  {Interactions}. \emph{Phys. Rev.} \textbf{1952}, \emph{85}, 338--353\relax
\mciteBstWouldAddEndPuncttrue
\mciteSetBstMidEndSepPunct{\mcitedefaultmidpunct}
{\mcitedefaultendpunct}{\mcitedefaultseppunct}\relax
\EndOfBibitem
\bibitem[Hummel(2015)]{hummel_2015}
Hummel,~F.~A. Density functional theory applied to liquid metals and the
  adjacent pair exchange correction to the random phase approximation. Ph.D.\
  thesis, University of Vienna, Vienna, 2015\relax
\mciteBstWouldAddEndPuncttrue
\mciteSetBstMidEndSepPunct{\mcitedefaultmidpunct}
{\mcitedefaultendpunct}{\mcitedefaultseppunct}\relax
\EndOfBibitem
\bibitem[Gruber \latin{et~al.}(2018)Gruber, Liao, Tsatsoulis, Hummel, and
  Gr\"uneis]{gruber_2018}
Gruber,~T.; Liao,~K.; Tsatsoulis,~T.; Hummel,~F.; Gr\"uneis,~A. Applying the
  Coupled-Cluster Ansatz to Solids and Surfaces in the Thermodynamic Limit.
  \emph{Phys. Rev. X} \textbf{2018}, \emph{8}, 021043\relax
\mciteBstWouldAddEndPuncttrue
\mciteSetBstMidEndSepPunct{\mcitedefaultmidpunct}
{\mcitedefaultendpunct}{\mcitedefaultseppunct}\relax
\EndOfBibitem
\end{mcitethebibliography}

\end{multicols}

\end{document}